\documentclass{article}
\usepackage{graphicx}
\usepackage{amsfonts}
\usepackage{amsmath}
\usepackage{amssymb}
\usepackage{url}
\usepackage{fancyhdr}
\usepackage{indentfirst}
\usepackage{enumerate}
\usepackage{csquotes}

\usepackage{dsfont}
\usepackage[colorlinks=true,citecolor=blue]{hyperref}
\usepackage{titlesec}
\usepackage{amsthm}
\usepackage{color}
\usepackage{natbib}
\usepackage{comment}
\usepackage{capt-of}
\usepackage{multirow}
\usepackage{subfig}
\def\d{\mathrm{d}}

\newcommand{\D}{\mathcal {D}}

\newcommand{\E}{\mathbb{E}}

\newcommand{\F}{\mathcal{F}}

\newcommand{\R}{\mathbb{R}}

\newcommand{\p}{\mathbb{P}}

\newcommand{\id}{\mathds{1}}

\newcommand{\btheta}{\boldsymbol{\theta}}
\newcommand{\mutheta}{\mu_{\boldsymbol{\theta}}}
\newcommand{\sigmatheta}{\sigma_{\boldsymbol{\theta}}}
\newcommand{\bpi}{\boldsymbol{\pi}}
\renewcommand{\(}{\left(}
\renewcommand{\)}{\right)}
\renewcommand{\[}{\left[}
\renewcommand{\]}{\right]}
\renewcommand{\ge}{\geqslant}

\renewcommand{\geq}{\geqslant}
\renewcommand{\leq}{\leqslant}
\renewcommand{\epsilon}{\varepsilon}

\renewcommand{\cdots}{\dots}

\theoremstyle{plain}
\newtheorem{theorem}{Theorem}

\newtheorem{proposition}{Proposition}
\theoremstyle{definition}
\newtheorem{definition}{Definition}

\theoremstyle{remark}
\newtheorem{remark}{Remark}

\newcommand{\cet}{\begin{center}}
\newcommand{\ecet}{\end{center}}

\allowdisplaybreaks

\usepackage{setspace}

\begin{document}
	
\title{
Behavioral Participating Insurance: Optimal Investment under Probability Distortion and Aspiration Constraints
} 

\author{
    Hao Liu\thanks{\scriptsize Department of Management Science and Engineering, Stanford University, USA. Email: \texttt{haoliu20@stanford.edu}}
	\and
	Yang Liu\thanks{\scriptsize 
    School of Science and Engineering, The Chinese University of Hong Kong (Shenzhen), China. Email: \texttt{yangliu16@cuhk.edu.cn}} 	
    \and
	Zhenyu Shen\thanks{\scriptsize Corresponding author. School of Science and Engineering, The Chinese University of Hong Kong (Shenzhen), China. Email: \texttt{zhenyushen@link.cuhk.edu.cn}} 
}

\date{}
\maketitle

\begin{abstract}

We study optimal investment for insurers managing participating (profit-sharing) contracts under probability distortion and probability benchmark (aspiration) constraints. The problem combines three theoretical complexities: (i) nonconcave effective utilities induced by embedded guarantees and surplus-sharing rules, (ii) probability weighting capturing behavioral aspects of long-horizon decisions, and (iii) aspiration-type constraints formalizing solvency requirements. Using quantile formulations and concavification techniques, we derive explicit closed-form solutions for optimal terminal wealth and trading strategies in both complete and incomplete Black-Scholes markets. Our utility class accommodates the piecewise hyperbolic absolute risk aversion (PHARA) family and covers nonconcavities arising naturally in insurance contexts. The framework reveals how probability distortion weakens lock-in behavior and induces time inconsistency: under inverse S-shaped distortions, insurers overestimate upside probabilities and increase risky investment relative to undistorted benchmarks. Asymptotic analysis and numerical illustrations demonstrate regime switches in optimal policies driven by regulatory thresholds and capital constraints. Our results extend the hope-fear-aspirations framework of He and Zhou (2016) and provide practical insights for managing insurance balance sheets under behavioral preferences and solvency constraints.


	
	
\end{abstract}

\section{Introduction}

Participating (profit-sharing) life insurance and annuity products remain a central pillar of long-horizon household saving and insurer asset-liability management. Their key features are guaranteed benefits combined with surplus participation rules, which create nonlinear payoffs and managerial incentives that are naturally studied through continuous-time surplus and portfolio choice models; see, e.g., \cite{BD1994,GK2007,BMS2010,BL2012,LSW2017,HLLM2019}. From the insurer's perspective, the terminal surplus is split between policyholders and shareholders through a contractual participation mechanism, which often induces \emph{nonconcave} effective objectives even when primitive preferences are concave (due to embedded guarantees, caps/floors, default boundaries, and convex compensation/participation components). This nonconcavity is now well documented in insurance-related investment problems with constraints and guarantee features; see, among others, \cite{CHN2019,NS2020,DZ2020,HLLM2020,HL2025}.

At the same time, a growing body of evidence supports the view that long-horizon investment and insurance decisions are shaped by \emph{probability distortion} (probability weighting) as in (cumulative) prospect theory; see \cite{KT1979}. Such distortions lead to rank-dependent criteria representable via (signed) Choquet integrals and are widely used to capture ``overweighting'' of tail events in insurance and finance; see \cite{WWW2020}. In dynamic portfolio choice, probability distortion breaks the tower property and thus destroys time consistency, making classical dynamic programming inapplicable. A key line of work in mathematical finance resolves this difficulty through \emph{quantile formulations} and Lagrange/duality arguments; see the foundational analytical treatment of cumulative prospect theory in \cite{HZ2011} and the ``hope-fear-aspirations'' framework of \cite{HZ2016}. Quantile formulation methods have been further clarified and systematized in, e.g., \cite{XZ2016,X2016,W2018,Bi2023}.

Motivated by insurance management with participating contracts, we study a continuous-time investment problem under (i) a general probability distortion and (ii) a broad class of potentially nonconcave utilities, additionally incorporating a \emph{probability benchmark (aspiration) constraint} of the form $ \mathbb{P}(X_T \ge L)\ge \alpha $. This benchmark formalizes a minimum-probability target that is natural in insurance contexts (e.g., a stylized solvency/guarantee requirement) and also aligns with the ``aspiration'' component emphasized in \cite{HZ2016}. Our analysis is carried out in both complete and incomplete Black-Scholes markets. The incomplete-market case is handled through the minimal market price of risk and a martingale/duality reduction to a terminal optimization over feasible wealth distributions; see \cite{Karatzas1991,KS1999} for classical incomplete-market duality.

The combination of (a) probability distortion, (b) nonconcave utilities induced by participating mechanisms, and (c) an aspiration-type probability benchmark yields an objective that is neither concave nor time-consistent. These features lead to two difficulties. First, probability distortion makes the objective time-inconsistent, so the dynamic programming approach is no longer directly applicable. Second, the participating mechanism produces a nonconcave effective utility, and the aspiration constraint imposes a probability requirement on terminal wealth. We therefore use a quantile formulation and a concave-envelope relaxation to characterize the optimal terminal payoff, identify the budget constraints in which the relaxed optimizer is attainable, and study how distortion and the aspiration constraint affect the insurer's risky investment.
    


\medskip
\noindent
\textbf{Our contributions.}
\begin{itemize}

    \item \textbf{Participating insurance contract management under distortion and benchmarks.}
    We develop a participating-contract model that leads to a PHARA-type utility and return to this model after establishing the general theory. We then study how probability distortion and the aspiration constraint affect the insurer's risky investment. The results show how these features change the classical undistorted strategy and how time inconsistency matters for the investment decisions made at the initial time both analytically and numerically.
    
    \item \textbf{A unified quantile/duality solution for nonconcave utilities under distortion.}
    Building on the quantile formulation approach of \cite{HZ2011,X2016,W2018} and the use of concave envelopes for nonconcave utilities, we obtain explicit formulas for the optimal terminal wealth. The utility class covers the PHARA family studied in \cite{LLMV2024} and includes the nonconcave utility functions generated by the participating-contract model. These formulas also recover several existing explicit solutions as special cases.

    \item \textbf{Time inconsistency and investment implications under probability distortion.}
    We examine how probability distortion changes the insurer's investment strategy over time. In the undistorted case, the optimal investment rule depends mainly on the remaining time to maturity. Under probability distortion, this property no longer holds: the strategy chosen at the initial time may differ from the one obtained by re-solving the problem at a later time. The numerical results illustrate this time inconsistency and show how it affects lock-in behavior and risky investment in participating-contract management.
    
\end{itemize}

\medskip
\noindent
\textbf{Related literature.}
Our work is most closely related to continuous-time portfolio choice under probability distortion and rank-dependent preferences. In the seminal paper, \cite{HZ2016} studies hope-fear-aspirations in a complete market and provides explicit solutions under a set of structural conditions; our analysis complements and extends this direction by (i) incorporating a probability benchmark constraint directly into the quantile formulation, (ii) accommodating a broad nonconcave utility family (including PHARA of \cite{LLMV2024}) induced by participating contracts, and (iii) treating incomplete markets via a minimal-kernel reduction.
\cite{Bi2023} develop a Lagrange-duality perspective for behavioral investment problems under distortion without the probability benchmark; our results can be viewed as adding a benchmark/aspiration layer together with feasibility/attainability guarantees and an insurance-motivated nonconcave utility class. Relatedly, \citet{BlanchetEtAl2025} study a time-inconsistent Merton problem under distributionally robust Bayesian control.
On the insurance side, optimal investment for participating contracts has been studied under classical expected utility and related objectives; see \cite{LSW2017,HLLM2020,LL2020} and references therein. More recently, \cite{BN2026} studies portfolio choice and contract design for participating insurance contracts under mismatch aversion. Nonconcave objectives and risk/solvency-type constraints appear naturally in guarantee products and insurance investment problems; see \cite{CHN2019,NS2020,DZ2020,LLZ2025}. Related behavioral investment problems with S-shaped utilities have also been studied in the pension context; for example, \cite{YJM2026} considers optimal investment with a path-dependent reference point. Our contribution is to connect these insurance-driven nonconcavities with rank-dependent distortion and an aspiration constraint in a unified explicit framework that applies in both complete and incomplete markets.

\medskip
\noindent
\textbf{Organization of the paper.}
Section~\ref{sec:motivation} introduces a motivating insurance-management example based on participating contracts and illustrates how the induced objective falls into the PHARA-type nonconcave family under probability distortion.
Section~\ref{sec:model} describes the market model (complete and incomplete cases), admissible strategies, and the distorted objective with the probability benchmark.
Section~\ref{sec:formulation} formulates the portfolio choice problem with the aspiration constraint. It also uses the labor-capital adjustment and the quantile formulation to transform the dynamic problem to a terminal optimization problem.
Section~\ref{formula} develops the unified quantile/duality solution and presents explicit formulas for the optimal terminal wealth and the associated precommitted strategy.
Section~\ref{sec:motivation revisit} returns to the participating-contract example and applies the general theory, followed by numerical illustrations. Economic implications and asymptotic behaviors are also discussed.
Section~\ref{sec:conclusion} concludes. Technical proofs are collected in the appendices.

\section{Participating Insurance Contract Management}\label{sec:motivation}
In this section, we present a motivating example of a defaultable participating contract, which comes from the insurance context (\cite{LSW2017,NS2020}). Let $X_T$ denote the fund value at some fixed time point $T$. The function $\Theta$ represents the payoff of the policyholder, which is a piece-wise affine function of $X_T$: 
$$
\begin{aligned}
\Theta(X_T) 
&= L_T^g + \delta (\eta X_T - L_T^g)_{+} - (L_T^g - X_T)_{+} \\
& = \left\{
\begin{aligned}
& X_T, && 0 \leq X_T < L_T^g;\\
& L_T^g, && L_T^g \leq X_T \leq \frac{L_T^g}{\eta};\\
& \delta \eta X_T + (1-\delta) L_T^g, && X_T > \frac{L_T^g}{\eta}.
\end{aligned}
\right.
\end{aligned} 
$$
Here, $L_0$ is the total contribution of the policyholder and also the initial liability of the insurer; $\eta$ is the initial liability-to-asset ratio of the insurer. Hence, the initial capital in the general account of the insurer is $x_0 = L_0/\eta > 0$. A minimum growth rate is denoted by $g \in (0, r)$. The guaranteed amount at maturity time $T$ is given by $L^g_T = L_0 e^{g T}$. 

The layer-feature of participating insurance contracts naturally fits with the benchmark-dependent nature of behavioral finance.
With the amount $\Theta(X_T)$ paid to the policyholder, the insurer retains a payoff as follows:
$$
\begin{aligned}
\Psi(X_T) &= X_T - \Theta(X_T) \\
& = \left\{
\begin{aligned}
& 0, && 0 \leq X_T < L_T^g;\\
& X_T - L_T^g, && L_T^g \leq X_T \leq \frac{L_T^g}{\eta};\\
&\left(1- \delta \eta\right) X_T - (1-\delta) L_T^g, && X_T > \frac{L_T^g}{\eta}.
\end{aligned}
\right.
\end{aligned} 
$$
The insurance interpretation of $\Psi(X_T)$ is clear. When the fund performs well, the insurer retains a large part of the surplus. When the fund value is in the intermediate region, the policyholder receives the guaranteed amount $L_T^g$, and the insurer keeps the remaining part of the fund value. When the fund suffers a severe loss and falls below $L_T^g$, the fund is not enough to cover the guarantee and the loss is borne by the policyholder. This layered payoff structure is also related to first-loss contracts in asset management; see \cite{HK2018}.
We consider a behavioral insurer with utility of the CPT type: 
$$
U_1(x) = 
\left\{
\begin{aligned}
& (x-B)^p, && x \geq B;\\
& -k (B-x)^q, && x < B,
\end{aligned}
\right.
$$
where $p,q \in (0,1)$, and $B$ is the benchmark of the insurer.
For simplicity of notation, we assume $p = q = \gamma$.
After composition, the effective utility function of the insurer is $U(X_T) = U_1\left(\Psi(X_T)\right) = U_1 (X_T - \Theta(X_T))$. Based on the values of $B$ and $\frac{1-\eta}{\eta}L_T^g$, we have the following three cases:

\noindent
Case 1: $0<B\leq \frac{1-\eta}{\eta}L_T^g$. The effective/composed utility is given by
\begin{equation}
U(X_T) = \left\{
\begin{aligned}
&-k B^q, && 0\leq X_T < L_T^g;\\
&-k\(B - X_T + L_T^g \)^q, && L_T^g \leq X_T \leq B + L_T^g;\\
&\(X_T - L_T^g - B\)^p, && B + L_T^g < X_T \leq \frac{L_T^g}{\eta};\\
&\(\(1 -\delta \eta\)X_T - \(1-\delta\)L_T^g - B \)^p, && X_T > \frac{L_T^g}{\eta}.
\end{aligned}
\right.
\end{equation}

\noindent
Case 2: $B> \frac{1-\eta}{\eta}L_T^g$. The composed utility is given by
\begin{equation}
U(X_T) = \left\{
\begin{aligned}
&-k B^q, && 0\leq X_T < L_T^g;\\
&-k\(B - X_T + L_T^g \)^q, && L_T^g \leq X_T \leq \frac{L_T^g}{\eta} ;\\
&-k\(B -\(1 - \delta\eta\)X_T + \(1-\delta\)L_T^g \)^q , && \frac{L_T^g}{\eta} < X_T \leq \frac{B +\(1-\delta\)L_T^g }{1-\delta\eta} ;\\
&\(\(1 -\delta \eta\)X_T - \(1-\delta\)L_T^g - B \)^p, && X_T > \frac{B +\(1-\delta\)L_T^g }{1-\delta\eta}.
\end{aligned}
\right.
\end{equation}

\noindent
Case 3: $B=0$. The composed utility is given by
\begin{equation}
U(X_T) = \left\{
\begin{aligned}
&0 && 0, \leq X_T < L_T^g;\\
&\(X_T - L_T^g\)^p, && L_T^g\leq X_T \leq \frac{L_T^g}{\eta};\\
&\(\(1-\delta\eta\)X_T - \(1-\delta\)L_T^g \)^p, && X_T > \frac{L_T^g}{\eta}.
\end{aligned}
\right.
\end{equation}
We see that the composed utilities are nonconcave in all three cases; some graphical illustrations are shown in Figure \ref{fig:three envelopes} later. \cite{Bi2023} study the utility maximization problem under probability distortion of settings including the third case. In a later context, we extend this analysis to a more general utility class, namely the piecewise HARA (PHARA) family of \cite{LLMV2024}. It is important to note, however, that the raw composed utilities $U=U_1\circ\Psi$ are not of the PHARA family in general. The PHARA structure enters through concavification: in the participating-contract context considered below, the relevant concave envelopes are piecewise HARA and hence fall into the PHARA family. We will therefore apply our general results to the concave envelopes, and later specify the conditions under which the resulting solution also solves the original nonconcave problem. We revisit this example in Section \ref{sec:motivation revisit} to apply the theoretical results and provide numerical illustrations.


\section{Model settings}\label{sec:model}

We consider a multi-dimensional Black-Scholes model. To obtain the explicit solution, all the market parameters are deterministic processes. The market consists of one risk-free asset and several risky assets. The risk-free asset has a deterministic return rate and no volatility, while the risky assets have higher expected return rates and positive volatilities. For a matrix $\mathbf{A}$, we denote by $\mathbf{A}^\top$ and $\mathbf{A}^{-1}$ its transpose and inverse (if exists). We denote the filtered probability space by $ \left( \Omega, \F_T, \{ \F_t \}_{0 \leq t \leq T}, \p \right)$. The filtration $\{ \F _t \}_{0 \leq t \leq T}$ is the one generated by a $q$-dimensional standard independent Brownian motion $\{ \mathbf{W}_t \}_{0 \leq t \leq T} = \{\(W_{1,t}, \ldots, W_{q,t}\)^\top \}_{0 \leq t \leq T}$ and further augmented by all $\p$-null sets. Let the deterministic process $\{ r_t \}_{0\leq t \leq T}$ denote the risk-free rate, where $\int_0^T\left|r_t\right|dt < \infty$. The risk-free asset $\{S_{0,t}\}_{0 \leq t \leq T}$ satisfies the ordinary differential equation
\begin{equation}\label{Bond}
\d S_{0,t} = r_t S_{0,t} \d t,\quad 0 \leq t \leq T.
\end{equation}
For the $m$ risky assets, we denote the return rate by a vector $\boldsymbol{\mu}_t := \(\mu_{1,t},\ldots, \mu_{m,t} \)^\top $ and the volatility by an $m \times q$ matrix $\boldsymbol{\sigma}_t = \{\sigma_{i,j,t}\}_{1 \leq i \leq m, 1 \leq j \leq q}$. We assume that $\boldsymbol{\sigma}_t \boldsymbol{\sigma}_t^\top$ is positive definite for all $t \in [0,T]$, and hence is invertible. Moreover, we assume that
\begin{equation*}
    \int_0^T\( \sum_{i = 1}^m\left|\mu_{i,t}\right| + \sum_{i = 1}^m\sum_{j = 1}^q \left| \sigma_{i,j,t}\right|^2\)\, \d t < \infty.
\end{equation*}

In Black-Scholes models, market completeness means that every contingent claim in the market can be replicated by a self-financing trading strategy, which is equivalent to the assumption that every risk can be hedged (i.e., $m=q$). As a study on the incomplete market, we consider that the risks may not be totally hedged, i.e., $m \leq q$. We assume that $\mu_{i,t} > r_t$ for $i = 1, \ldots, m, \; t \in \[0, T\]$. The evolution of the $i$-th risky asset follows the geometric Brownian motion which satisfies the stochastic differential equation:
\begin{equation}\label{Stock}
\d S_{i,t} = \mu_{i,t} S_{i,t} \d t + S_{i,t} \boldsymbol{\sigma}_{i,t}^\top \d \mathbf{W}_t, \quad i = 1, \ldots, m,
\end{equation}
where $\boldsymbol{\sigma}_{i,t}$ denotes the $i$th row of $\boldsymbol{\sigma}_t$. Letting $\mathbf{1}_m := (1, \ldots, 1)^\top \in \R^m$, we define
$$\Theta = \left\{\boldsymbol{\theta}': \boldsymbol{\sigma}_t \boldsymbol{\theta}'_t =  \boldsymbol{\mu}_t - r_t \mathbf{1}_m \text{ almost surely for almost every }t \in [0,T] \text{ with }\exp\left(\int_0^T \left\Vert\boldsymbol{\theta}'_t\right\Vert_2^2dt\right) < \infty\right\},$$
which is assumed to be nonempty, convex, and $\mathcal{F}_t$-progressively measurable. Here we define $\| \boldsymbol{\theta}'_t \|_2 := \(\sum_{i = 1}^q |\theta'_{i,t}|^2\)^{\frac12}$.
Moreover, we define
\begin{equation}\label{eq:chosen theta}
\boldsymbol{\theta}_t := \boldsymbol{\sigma}_t^\top \( \boldsymbol{\sigma}_t \boldsymbol{\sigma}_t^\top \)^{-1} \( \boldsymbol{\mu}_t - r_t \mathbf{1}_m \),
\end{equation}
which represents the vector of the minimal market price of risk. Note that when $m = q$ (in complete market), $\boldsymbol{\theta}_t$ in Equation \eqref{eq:chosen theta} reduces to $\boldsymbol{\theta}_{0,t}$ given by
\begin{equation}\label{eq:complete theta} 
\boldsymbol{\theta}_{0,t} := \boldsymbol{\sigma}_t^{-1} \( \boldsymbol{\mu}_t - r_t \mathbf{1}_m \),
\end{equation}
which is the unique element in $\Theta$.
\begin{remark}
In the incomplete Black-Scholes model, the market price of risk $\{\boldsymbol{\Tilde{\theta}}_t \}_{0\leq t\leq T}$ can be any solution of the linear system:
\begin{equation}\label{eq:price of risk}
\boldsymbol{\sigma}_t \boldsymbol{\Tilde{\theta}}_t = \boldsymbol{\mu}_t - r_t \mathbf{1}_m.
\end{equation}
Hence, if $m < q$, $\boldsymbol{\Tilde{\theta}}_t$ is not unique. However, if all the market parametric processes are deterministic, then the least squared problem 
\begin{equation}\label{convexCone}
    \arg \min_{\boldsymbol{\Tilde{\theta}_t} \in \Theta } \left\Vert\boldsymbol{\Tilde{\theta}}_t\right\Vert_2^2
\end{equation}has a unique solution $\hat{\boldsymbol{\theta}}_t = \boldsymbol{\theta}_t$.
\end{remark}
Next, for each parametric process $\boldsymbol{\Tilde{\theta}}_t \in \Theta$, we define the pricing kernel process $\{ \xi_{\boldsymbol{\Tilde{\theta}},t} \}_{0 \leq t \leq T}$ as follows:
\begin{equation}\label{pricingkernel}
\xi_{\boldsymbol{\Tilde{\theta}},t} := \exp \left\{ - \int_0^t \( r_s + \frac12 \| \boldsymbol{\Tilde{\theta}}_s \|_2^2 \)\, \d s - \int_0^t \boldsymbol{\Tilde{\theta}}_s^\top \, \d \mathbf{W}_s\right\}, \quad 0 \leq t \leq T.
\end{equation}
Since $r_t$ is integrable, we have $0 < \xi_{\boldsymbol{\Tilde{\theta}},T} < \infty$ almost surely and $0 < \mathbb{E}[\xi_{\boldsymbol{\Tilde{\theta}},T}] < \infty$. Moreover, if we denote $\xi_{\boldsymbol{\Tilde{\theta}}}: = \xi_{\boldsymbol{\Tilde{\theta}},T}$ and assume it is atomless (i.e. has a continuous distribution), then $\xi_{\boldsymbol{\Tilde{\theta}}}$ is log-normally distributed with cumulative distribution function 
$$F_{\xi_{\boldsymbol{\Tilde{\theta}}}}(x) = \Phi\left( \frac{\log x - \mu_{\boldsymbol{\Tilde{\theta}}}}{\sigma_{\boldsymbol{\Tilde{\theta}}}} \right),$$
where $\Phi$ is the standard normal cumulative distribution function, 
$\mu_{\boldsymbol{\Tilde{\theta}}} = -\int_0^T \left(r_s + \frac{1}{2}\left\Vert\boldsymbol{\Tilde{\theta}}_s\right\Vert_2^2\right)\d s,$
 and
$\sigma_{\boldsymbol{\Tilde{\theta}}} = \sqrt{\int_0^T  \left\Vert\boldsymbol{\Tilde{\theta}}_s\right\Vert_2^2\d s}.$ 
We denote by $\{\pi_{i,t}\}_{0 \leq t \leq T}$ the amount of money invested in $i$th risky asset $S_{i}$ at time $t$. The wealth process $\{X_t\}_{0 \leq t \leq T}$ is uniquely determined by the investment process $\{\boldsymbol{\pi}_t = \(\pi_{1,t}, \ldots, \pi_{m,t} \)^\top \}_ {0 \leq t \leq T}$ and an initial value $x_0 > 0$:
\begin{equation}\label{wealthprocess}
\d X_t = \( r_t X_t + \boldsymbol{\pi}_t^\top \(  \boldsymbol{\mu}_t - r_t \mathbf{1}_m\)\) \d t + \boldsymbol{\pi}_t^\top \boldsymbol{\sigma}_t \d \mathbf{W}_t + C_t \d t, \quad X_0 = x_0,
\end{equation}
where $\{C_t\}_{0\leq t\leq T}$ is a deterministic process denoting the rate of insurance fee received; see \cite{HLLM2019}.
A portfolio $\boldsymbol{\pi} = \{ \boldsymbol{\pi}_t \}_{0 \leq t \leq T}$ is called admissible if and only if $\boldsymbol{\pi}$ is an $\{ \mathcal{F}_t \}_{0 \leq t \leq T}$ -progressively measurable $\R^m$-valued process, $\int_0^T \|\boldsymbol{\sigma}_t^\top\boldsymbol{\pi}_t\|_2^2 \, \d t < \infty$, and $\int_0^T \left| \boldsymbol{\pi}_t^\top \left(\boldsymbol{\mu}_t - r_t\mathbf{1}_m \right) \right| \d t < \infty$ almost surely. We define $\mathcal{V}$ as the collection of all admissible portfolios.

In our context, we consider a utility function with enough generality, the PHARA utility function. 

\begin{definition}[PHARA utility]
	\label{def_PHARA}
	Define the function
	$\tilde{U}: \R \to \R$ (with relative risk aversion parameter $R \in [0, \infty]$, benchmark level $A \in \R$, partition point $\hat{x} \in \R$, utility value $u \in \R$, slope $\gamma \in (0, \infty)$, absolute risk aversion parameter $\alpha \in (0, \infty)$) as
	\begin{equation}%
		\label{eq_def_U_tilde}
		\begin{aligned}
			\tilde{U}(x; R, A, \hat{x}, u, \gamma, \alpha) 
			:=
			\left\{
			\begin{aligned}
				& \gamma ( x-\hat{x} ) + u, && \text{ if } R = 0;\\
				& \gamma (\hat{x}-A) \log\left(\frac{x-A}{\hat{x}-A}\right) + u, && \text{ if } R = 1, \hat{x} \neq A;\\
				& \gamma \frac{\hat{x}-A}{1-R} \left( \left(\frac{x-A}{\hat{x}-A}\right)^{1-R}-1 \right) + u, && \text{ if } R \in (0, 1) \cup (1, \infty), \hat{x} \neq A;\\
				& 
				-\frac{\gamma}{\alpha} \left( e^{-\alpha(x-\hat{x})} -1 \right) +u, && \text{ if } R = \infty, A = -\infty, \alpha \in (0, \infty). 
			\end{aligned}
			\right.
		\end{aligned}
	\end{equation}
	Set the effective domain $\mathcal{D} := [a_0, \infty)$ or $(a_0, \infty)$, where $a_0  := \inf\{a \in \R | U(a) > -\infty\}$. 
	A function $U: \R \to \R\cup\{-\infty\}$ is a \textit{piecewise HARA utility} if and only if there exists a partition 
	$\{a_k\}_{k = 0}^{n+1}$ and a family of parameter pairs $\{(R_k, A_k, \alpha_k)\}_{k=0}^n$ (note that it is no need to specify $A_k$ for $k$ satisfying $R_k = 0$, i.e., on the linear part) such that 
	\begin{enumerate}[(i)]
		\item $n \geq 0$, $a_0 < a_1 < \cdots < a_{n+1}$, $a_1, \cdots, a_n \in \R$, $a_{n+1} = \infty$;
		\item $U$ is increasing and continuous on $\mathcal{D}$ and $U(x) = -\infty$ for $x \notin \mathcal{D}$;
		\item 
		\begin{enumerate}[(a)]
			\item
			If $n = 0$ and $\mathcal{D} = (a_0, \infty)$, or $n =0$ and $\mathcal{D} = [a_0, \infty)$ and $A_0 = a_0$, then 
			$U(x) = \tilde{U}^0(x; R, A, u, \gamma, \alpha)$ for any $x \in (a_0, \infty)$, where 
			\begin{equation}%
				\label{eq_def_U_tilde_0}
				\begin{aligned}
					\tilde{U}^0(x; R, A, u, \gamma, \alpha) 
					:=
					\left\{
					\begin{aligned}
						& \gamma x + u, && \text{ if } R = 0;\\
						& \gamma \log (x-A) + u, && \text{ if } R = 1;\\
						& \frac{\gamma}{1-R}(x-A)^{1-R} + u, && \text{ if } R \in (0, 1) \cup (1, \infty);\\
						& 
						-\frac{\gamma}{\alpha} \left( e^{-\alpha x} -1 \right) +u, && \text{ if } R = \infty, A = -\infty, \alpha \in (0, \infty). 
					\end{aligned}
					\right.
				\end{aligned}
			\end{equation}
			\item 
			If $n = 0$ and $\mathcal{D} = [a_0, \infty)$, then $U(x) = \tilde{U}(x; R_0, A_0, a_0, U(a_0), U'(a_0+), \alpha_0)$ for any $x \in [a_0, \infty)$.
			\item 
			If $n \geq 1$, for $k = 0$, $U(x) = \tilde{U}(x; R_0, A_0, a_1, U(a_1), U'(a_1-), \alpha_0)$ for any $x \in (a_0, a_1)$; 
			for any $k\in\{1, 2, \cdots, n\}$, $U(x) = \tilde{U}(x; R_k, A_k, a_k, U(a_k), U'(a_k+), \alpha_k)$ for any $x \in (a_k, a_{k+1})$.
		\end{enumerate}
		\item For $n$, $R_n \neq 0$ (i.e., $U$ is not linear on the last interval $(a_n, \infty)$). 
	\end{enumerate}
\end{definition}
Finally, for simplicity, we use the notation: $\gamma_k^+ = U'(a_k+)$, $\gamma_k^- = U'(a_k-)$, where $\gamma_0^- = \infty$ and $\gamma_{n+1}^- = 0$.

To solve the optimization problem explicitly, we must specify the functional form of the probability distortion $w(\cdot)$. While our later theoretical framework accommodates any inverse S-shaped distortion satisfying standard regularity conditions, deriving closed-form solutions for the optimal terminal wealth and trading strategies requires a parametric family that preserves analytical tractability under the quantile transformation. Specifically, the chosen family must allow the composition of the distortion function with the Gaussian pricing kernel to remain within a manageable functional class.

In line with \cite{HZ2016} on the ``hope-fear-aspirations'' framework, we adopt a flexible parametric class of distortions based on normal shifts.
Parametrized by $(a,b,\tilde{z})$, this class of distortion functions is defined by
\begin{equation}\label{distortion}
w(z) =
\begin{cases}
ke^{(a+b)\Phi^{-1}(\bar{z}) + \frac{a^2}{2}} \Phi(\Phi^{-1}(z) + a), & z \leq \bar{z}, \\
A + ke^{\frac{b^2}{2}} \Phi(\Phi^{-1}(z) - b), & z \geq \bar{z},
\end{cases}
\end{equation}
where $a,b \geq 0$, $A = 1 - ke^{\frac{b^2}{2}}$, and
$$k = \frac{1}{e^{\frac{b^2}{2}} \Phi(-\Phi^{-1}(\bar{z}) + b) + e^{(a+b)\Phi^{-1}(\bar{z}) + \frac{a^2}{2}} \Phi(\Phi^{-1}(\bar{z}) + a)}.$$
This family is particularly well-suited for our Black-Scholes setting because it maintains the Gaussian structure of the pricing kernel, thereby enabling the explicit evaluation of the integrals arising in the martingale duality approach. Moreover, by adjusting the parameters, this class can capture the full spectrum of behavioral attitudes, from the overweighting of tail events (hope and fear) to the standard expected utility benchmark.

\section{Optimization problems}\label{sec:formulation}
\subsection{Terminal wealth optimization}
Before presenting the problem that will be solved in this paper, we firstly review a classical optimization problem in financial mathematics. With the model setting as above, we want to solve the problem 
\begin{equation}\label{prob-main}
	\begin{aligned}
		& \sup_{\boldsymbol{\pi} \in \mathcal{V}} \E[ U(X_T)],
	\end{aligned}
\end{equation}
where $U$ is called a utility function defined as above and $\mathcal{V}$ is the collection of all the admissible portfolios. 

In our context, the utility function $U$ may not be concave and cause difficulties in the solving procedure. A highly related concept is the concave envelope. Let $\D \subseteq \R$ be a convex set. Denote a continuous function by $U: \D \to \R$, where the domain of $U$ is denoted by $\text{dom } U = \D$. The concave envelope of $U$ is defined as the smallest 
continuous concave function larger than $U$. That is, for $x \in \D$, 
\begin{equation}\label{eq:conv_dominating}
	\hat{U}(x) := \inf\{h(x): h \text{ maps $\D$ to $\R$, $h$ is a concave and continuous function on $\D$ and } h \geq U\}.
\end{equation}
Hence, for a continuous utility function $U$, the concave envelope $\hat{U}$ is concave, 
continuous and larger than $U$, and $\hat{U}$ is the smallest function satisfying these three conditions.

\begin{remark}
Under mild assumptions, the existence of Lagrangian multiplier and well-posedness of Problem \eqref{prob-main} are always true.
The value of $x_0$ only changes the value of the optimal control but not the structure. This will become a specific problem in the study of the general portfolio choice problem.
\end{remark}

\begin{remark}
    The concave envelope plays an important role in portfolio selection problems with nonconcave utilities. Indeed, as the literature of concavification principle indicates (e.g., \cite{LL2020}), since $\xi_{\boldsymbol{\theta}}$ has a continuous distribution, Problem \eqref{prob-main} has the same optimal solution as the following problem with the utility $U$ replaced by the concave envelope $\hat{U}$:
\begin{equation}\label{prob:concavification}
	\max_{\boldsymbol{\pi} \in \mathcal{V}} \E\left[ \hat{U}(X_T) \right].
\end{equation}
Thus, we can study $\hat{U}$ instead of $U$ and later obtain the same optimal portfolio for both Problems \eqref{prob-main} and \eqref{prob:concavification}. However, when the optimization problem becomes general, this equivalence is not automatic.
\end{remark}

Further, we consider a parametric family of distortions proposed by \cite{HZ2016} to get the explicit formula of optimal controls. The same computational methods can be applied to other popular distortion functions with numerical simulations. 

In an incomplete market ($m < q$), under our assumption, the computation is essentially the same, except changing the parametric process from $\boldsymbol{\theta}_{0}$ to $\boldsymbol{\theta}$. The kernel process with this $\boldsymbol{\theta}$ is often called the minimal pricing kernel. The proof of this optimality can be viewed as a special case of the general problem; we defer this justification to Section \ref{formula}.

\subsection{Labor capital}\label{sec:labor capital}
The SDE \eqref{wealthprocess} lacks the self-financing property due to the term $C_t$ and we use a labor capital technique to recover the property (see \cite{HH2012} and \cite{HLLM2019}).
We define $H(t)$ as the labor capital per unit contribution at time $t$:
\begin{equation*}
H(t) = \int_t^T e^{-\int_t^s r_u \d u }C_s \d s.
\end{equation*}
Then $H(t)$ represents the value of future deterministic income at time $t$. The total wealth $Y_t$ at time $t$ satisfies the following SDE:
\begin{equation*}
    \d Y_t = \(r_t Y_t + \boldsymbol{\pi}_t^\top\(\boldsymbol{\mu}_t - r_t \mathbf{1}_m\) \) \d t + \boldsymbol{\sigma}_t^\top \boldsymbol{\pi}_t \d \mathbf{W}_t,\quad Y_0 = x_0 + H(0).
\end{equation*}
Using the labor capital technique (\cite{HLLM2019}), we see that Problem \eqref{prob-main} can be transformed into the following terminal optimization problem:
\begin{equation}\label{eq:auxiliary problem}
    \sup_{Y_T \in \mathcal{V}\(Y_0\)} \E \[U\(Y_T\) \],
\end{equation}
where 
\begin{equation*}
    \mathcal{V} \(Y_0\) := \left\{Y_T: Y_T \text{ is $\F_T$ measurable and }\E[\xi_{\btheta}Y_T ] =Y_0 \right\}.
\end{equation*}

\subsection{General portfolio choice problem with probability benchmarks}\label{genearl}
From \cite{HZ2016}, we study the distorted-probability optimization problem:
\begin{equation}\label{eq: original}
\begin{aligned}
\sup_{\boldsymbol{\pi} \in \mathcal{V}[0,T]} \quad &\int_0^\infty U(x) \d\left(1-w(1 - F_{X_T}(x))\right) \\
\text{subject to} \quad &\d X_t = \( r_t X_t + \boldsymbol{\pi}_t^\top \(  \boldsymbol{\mu}_t - r_t \mathbf{1}_m\)\) \d t + \boldsymbol{\pi}_t^\top \boldsymbol{\sigma}_t \d \mathbf{W}_t + C_t \d t, \quad t \in [0, T]; \\
    &X_0 = x_0, \quad \p(X_T \geq L) \geq \alpha, 
\end{aligned}
\end{equation}
where $0<\alpha <1$, $L \geq 0$, and $w: [0,1] \to [0,1]$ is inverse $S$-shaped, strictly increasing, and continuously differentiable with $w(0) = 0$ and $w(1) = 1$. Recall that distortion functions defined in Section \ref{sec:model} satisfy these conditions.
Note that when $w$ is the identity function and $L=0$, Problem (\ref{eq: original}) reduces to Problem (\ref{prob-main}). Indeed, \cite{HZ2016} studies the same problem under a complete market with the assumption that $U$ is strictly concave. Here, we only assume that $\hat{U}$ is of the PHARA family. This assumption is weak and general enough to cover almost all cases. The existence of the distortion function $w$ makes the problem time-inconsistent, which destroys the use of dynamic programming approach and hence we turn to the martingale method. For the ease of notations, we first consider the complete market case, where $m = q$. Since in this case $\boldsymbol{\theta}_t = \boldsymbol{\theta}_{0,t}$, we still use $\{\boldsymbol{\theta}_t\}_{0\leq t\leq T}$ for notation simplicity.

In the spirit of the martingale method in incomplete markets (\cite{Karatzas1991}), it suffices to solve the terminal wealth optimization problem:
\begin{equation}\label{eq:new}
\begin{aligned}
\max_{Y_T \in \mathcal{\Tilde{F}}_T} \quad &\int_0^\infty U(x) \d\left(1-w(1 - F_{Y_T}(x))\right) \\
\text{subject to} \quad & \E[\xi_{\boldsymbol{\theta}} Y_T]  = x_0 + H(0) , \quad \p(Y_T \geq L) \geq \alpha.
\end{aligned}
\end{equation}
In order to solve Problem (\ref{eq:new}), we use a quantile formulation method (\cite{HZ2011}) and transform it into the following problem:
\begin{equation}\label{eq:QF}
\begin{aligned}
    &\sup_{G \in \mathcal{G}_L} \int_{0}^{1} U(G(x))w'(1-x) \, \d x \\
    \text{subject to} \quad &\int_0^1 G(x)F_{\xi_{\boldsymbol{\theta}}}^{-1}(1-x) \, \d x = x_0 + H(0) , 
    \quad G((1-\alpha)+) \geq L,
\end{aligned}
\end{equation}
where
$$\mathcal{G}_L = \{G: (0,1) \to \mathbb{R}, \text{ increasing and right continuous with left limits, }G((1-\alpha)+)\geq L\}.$$


Instead of using the Lagrange multiplier method to solve Problem (\ref{eq:QF}) directly, we make a further transformation and establish equivalence between solutions of these problems. To be more precise, we apply the change of variable method similarly in \cite{X2016} to define
\begin{equation*}
    y = 1 - w^{-1}(1-x).
\end{equation*}
Since $\d x = w'(1-y)\d y$, the objective function becomes
\begin{equation*}
    \int_{0}^{1} U(G(y))w'(1-y) \, \d y =  \int_{0}^{1} U(Q(y)) \, \mathrm{d}y,
\end{equation*}
where
$$
Q(x) = G\left(1-w^{-1}(1-x)\right).
$$
Under the same transformation, the budget constraint can be rewritten by introducing
$$\varphi(x) = -\int_0^{w^{-1}(1-x)}F_{\xi_{\boldsymbol{\theta}}}^{-1}\left(z\right)\d z, \quad x \in [0,1],$$
whose derivative satisfies
\begin{equation*}
    \varphi'(x) = \frac{F_{\xi_{\btheta}}^{-1}\(w^{-1}(1-x)\)}{w'\(w^{-1}\(1-x\)\)}.
\end{equation*}
The aspiration constraint is transformed in the same way. The change of variable $x=1-w(1-y)$ gives $G((1-\alpha)+)=Q((1-w(\alpha))+).$ Since $Q$ is right-continuous, this is equal to $Q(1-w(\alpha))$. Hence the aspiration constraint becomes $Q(1-w(\alpha))\ge L$. Hence, Problem \eqref{eq:QF} can be transformed to:
\begin{equation}\label{eq:change}
    \begin{aligned}
    &\sup_{Q \in \mathcal{G}} \int_{0}^{1} U(Q(x)) \, \mathrm{d}x \\
    \text{subject to} \quad &\int_0^1 Q(x)\varphi'(x) \, \mathrm{d}x = x_0+H(0), 
    \quad Q(1-w(\alpha)) \geq L.
    \end{aligned}
\end{equation}

\section{A unified formula for solution with distortion}\label{formula}
We first present the notation. Let $\hat{\varphi}$ denote the concave envelope of $\varphi$.
Define $c = \inf\{x>0: \varphi(x) = \hat{\varphi}(x)\}$, and 
$$\Lambda:= \left\{0<  \lambda < \infty : \int_{0}^{1} f\left(\lambda \hat{\varphi}'(x +)\right)\varphi'(x) \, \d x < +\infty \right\},$$
    where $f: \mathbb{R} \to [0,\infty]$ is given by
    \begin{align*}
	f(x) =  \sum_{k=0}^n \biggr\{&a_k \id_{\left\{x \in (\gamma_k^{+}, \gamma_k^{-})\right\}}\\
	+&  \left( A_k + \left(\frac{\gamma_k^{+}}{x}\right)^{\!R_k^{-1}}\!\!(a_k - A_k) \right) \id_{\left\{x \in \left(\gamma_{k+1}^{-}, \gamma_k^{+}\right)\right\}} \id_{\{ R_k \neq 0,\infty \}}\\ 
	+& \left(a_k + \frac{1}{\alpha_k} \log\left(\frac{\gamma_k^{+}}{x}\right)\right) \id_{\left\{x \in \left(\gamma_{k+1}^{-}, \gamma_k^{+}\right)\right\}} \id_{\left\{ R_k = \infty, A_k = -\infty, \alpha_k > 0 \right\}} \biggr\}.
\end{align*}
For $k = 0, \ldots, n-1$, define 
    \begin{equation*}
\lambda_k=
\left\{
\begin{aligned}
     &\frac{\gamma_k^+}{\varphi'(c)}, &&R_k = 0;\\
    &0, &&R_k \neq 0.
\end{aligned}
\right.
\end{equation*}
Further, define $\lambda_n := U'(L)/\varphi'(c)$ and
$\Lambda_n =\{\lambda_k: k = 0 ,\ldots, n \text{ such that }\lambda_k > 0\}$. Therefore, $\left|\Lambda_n\right| \leq n$ since the last part is nonlinear. For $\lambda_k \in \Lambda_n$, define
$I_k = (x_{L,k}, x_{R,k})$, where
\begin{equation*}
\begin{aligned}
    x_{L,k} = \int_0^1 Q_L(\lambda_k,x)\varphi'(x)\d x,\quad
    x_{R,k} = \int_0^1 Q_R(\lambda_k,x)\varphi'(x)\d x,
\end{aligned}
\end{equation*}
where
\begin{equation*}
\begin{aligned}
    Q_L(\lambda, z) &:= I^L\(\lambda \hat{\varphi}'(z) \)\id_{\{z\leq 1-w(\alpha) \}} + \[I^L\(\lambda \hat{\varphi}'(z)\) \vee L \]\id_{\{z > 1-w(\alpha)\} },\\
    Q_R(\lambda, z) &:= I^R\(\lambda \hat{\varphi}'(z) \)\id_{\{z\leq 1-w(\alpha) \}} + \[I^R\(\lambda \hat{\varphi}'(z)\) \vee L \]\id_{\{z > 1-w(\alpha)\} }.
\end{aligned}
\end{equation*}
We assume that $\Lambda$ is nonempty since the emptiness of $\Lambda$ implies that Problem \eqref{eq:change} is ill-posed from Theorem 3.1 in \cite{JXZ2007}.



Given the quantile formulation, here if we choose $g(x) = G^*(1-F_{\xi_{\boldsymbol{\theta}}}(x)) $, then $Y^*$ is replicable by the initial wealth $x_0+H(0)$. This implies that under the deterministic parameter process condition and assumptions on the set $\Theta$, we can essentially use the same argument as in the complete market to derive the optimal solutions for the incomplete market.

Recall that 
$$\Theta = \left\{\boldsymbol{\theta}': \boldsymbol{\sigma}_t \boldsymbol{\theta}'_t =  \boldsymbol{\mu}_t - r_t \mathbf{1}_m \text{ almost surely for almost every }t \in [0,T] \text{ with }\exp\left(\int_0^T \left\Vert\boldsymbol{\theta}'_t\right\Vert_2^2dt\right) < \infty\right\},$$
where Proposition 3 in \cite{LiuShen2024} shows that the same conditions must hold for the usual expected terminal wealth maximization problem in the incomplete market to have the optimal controls computed from the minimal pricing kernel. According to Theorem 9.3 in \cite{Karatzas1991}, this condition is equivalent to the replicability of the terminal wealth with respect to the same initial wealth. 

Above arguments show another perspective that quantile formulation problems generalize the classical portfolio optimization problem. 
Before presenting our main results, we give some equations that will be used repeatedly. First, note that
\begin{equation*}
\xi_{\btheta}= \xi_t \exp\left\{-\int_t^T\left(r_s+\frac{\|\btheta_s\|_2^2}{2}\right)\d s-N\sqrt{\int_t^T\|\btheta_s\|_2^2\d s}\right\},
\end{equation*}
where
\begin{equation*}
N:=\frac{\int_t^T\btheta_s^\top\d W_s}{\sqrt{\int_t^T\|\btheta_s\|_2^2\d s}}
\end{equation*}
is a standard normal random variable conditional on $\F_t$.
For $t \in [0,T)$ and  $y\in [0,1]$, define the functions
\begin{equation}
\begin{aligned}
    d(t,y) &:= -\frac{1}{\sqrt{\int_t^T \|\btheta_s\|_2^2\,\d s}}\left\{\log\(\frac{F_{\xi_{\boldsymbol{\theta}}}^{-1}\(w^{-1}\(1-y\) \) }{\xi_t} \)+\int_t^T\(r_s - \frac{\|\btheta_s\|_2^2}{2}\)\,\d s \right\},\\
    \hat{d}(t,y) &:= d(t,y) - \sqrt{\int_t^T \|\btheta_s\|_2^2\,\d s},\\
    d_k(t,y) &:= d(t,y) -c_k\sqrt{\int_t^T \|\btheta_s\|_2^2\,\d s},\\
    \hat{d}_k(t,y) &:= d_k(t,y) -c_k\sqrt{\int_t^T \|\btheta_s\|_2^2\,\d s},\\
    B_{t,k} &:= \exp \left\{\frac{a\mutheta}{R_k \sigmatheta }+c_k\int_t^T\(r_s+\frac{(c_k-1)\|\btheta_s\|_2^2}{2}\)\,\d s \right\},
\end{aligned}
\end{equation}
where $c_k := \frac{\sigmatheta+a}{R_k\sigmatheta}$.
Also, we define the generalized inverse function of $\(\hat{\varphi}\)'$ by
\begin{equation}
    I (y) := \inf \left\{x \in [0,1]| \(\hat{\varphi}\)'(x) \leq y  \right\}.
\end{equation}
Finally, we define the following functions, which are essentially probability weights:
\begin{align}
    P\(t, \lambda, u,v\) &:= \Phi\(d\(t, u\)\vee d\(t, I\(\frac{U'\(L\)}{\lambda}\) \)\vee d\(t, 1-w\(\alpha\) \) \)\nonumber\\
    &\quad- \Phi\(d\(t,v\)\vee d\(t, I\(\frac{U'\(L\)}{\lambda}\) \)\vee d\(t, 1-w\(\alpha\) \) \)\nonumber\\
    &\quad + \Phi\(d\(t, u\)\wedge d\(t, 1-w(\alpha) \) \) - \Phi\(d\(t,v\)\wedge d\(t, 1-w(\alpha)\) \),\nonumber\\
    P_k\(t, \lambda, u,v\) &:= \Phi\(d_k\(t,u\)\vee d_k\(t, I\(\frac{U'\(L\)}{\lambda}\) \)\vee d_k\(t, 1-w\(\alpha\) \) \)\nonumber\\
    &\quad - \Phi\(d_k\(t,v\)\vee d_k\(t, I\(\frac{U'\(L\)}{\lambda}\) \)\vee d_k\(t, 1-w\(\alpha\) \) \)\nonumber\\
    &\quad + \Phi\(d_k\(t, u\)\wedge d_k\(t, 1-w(\alpha) \) \) - \Phi\(d_k\(t,v\)\wedge d_k\(t, 1-w(\alpha)\) \),\nonumber\\
    Q\(t, \lambda, u,v\) &:= \Phi'\(d\(t,u\)\vee d\(t, I\(\frac{U'\(L\)}{\lambda}\)\) \vee d\(t, 1-w\(\alpha\) \)\)\nonumber\\
    &\quad- \Phi'\(d\(t,v\)\vee d\(t,I\(\frac{U'\(L\)}{\lambda}\)\)\vee d\(t, 1-w\(\alpha\) \) \)\nonumber\\
    &\quad + \Phi'\(d\(t,u\)\wedge d\(t, 1-w(\alpha)\) \) - \Phi'\(d\(t,v\)\wedge d\(t,1-w(\alpha)\) \),\nonumber\\
    Q_k\(t,\lambda, u,v\) &:= \Phi'\(d_k\(t,u\)\vee d_k\(t, I\(\frac{U'\(L\)}{\lambda}\)\)\vee d_k\(t, 1-w\(\alpha\) \) \)\nonumber\\
    &\quad - \Phi'\(d_k\(t,v\)\vee d_k\(t,I\(\frac{U'\(L\)}{\lambda}\)\)\vee d_k\(t, 1-w\(\alpha\) \) \)\nonumber\\
    &\quad + \Phi'\(d_k\(t,u\)\wedge d_k\(t, 1-w(\alpha)\) \) - \Phi'\(d_k\(t,v\)\wedge d_k\(t,1-w(\alpha)\) \).
\end{align} 
The following proposition and theorem summarize the optimal terminal wealth, the optimal wealth process, and the optimal controls obtained by the martingale and duality method.

\begin{proposition}\label{prop:optimal wealth process}
Suppose that the concave envelope $U^{**}$ has the form in Definition \ref{def_PHARA} and $R_k \in [0, \infty]$ for each $k \in \{0,1,\cdots, n\}$. For fixed $\lambda>0$, we define the following processes:
\begin{align}\label{eq:wealth process basic}
    X_{t,k,\lambda}^D &:= e^{-\int_t^T r_s \,\d s}a_k P\(t,\lambda, I\(\frac{\gamma_k^+}{\lambda} \), I\(\frac{\gamma_k^-}{\lambda} \) \),\nonumber\\
    X_{t,k,\lambda}^P &:= e^{-\int_t^T r_s \,\d s}\left\{A_k P\(t,\lambda, I\(\frac{\gamma_{k+1}^-}{\lambda} \), I\(\frac{\gamma_k^+}{\lambda} \) \) \right.\nonumber\\
    &\quad \left.+\(a_{k}-A_{k} \)B_{t,k}\(\frac{\tilde{k}\gamma_{k}^+}{\lambda} \)^{R_{k}^{-1}} \xi_t^{-c_{k}}
    P_k\(t,\lambda,I\(\frac{\gamma_{k+1}^-}{\lambda} \), I\(\frac{\gamma_k^+}{\lambda} \) \) \right\}\id_{\{R_k \neq 0, \infty \}}\nonumber,\\
    X_{t,k,\lambda}^E 
    &=e^{-\int_t^T r_s \,\d s}\left\{a_k P\(t,\lambda, I\(\frac{\gamma_{k+1}^-}{\lambda}\), I\(\frac{\gamma_k^+}{\lambda} \) \) \right.\nonumber\\
    &\quad\quad + \frac{1}{\alpha_k}\left\{\(\log\( \frac{\tilde{k} \gamma_{k}^+ }{\lambda \xi_t^{1+\frac{a}{\sigmatheta}}} \)+\frac{a\mutheta }{\sigmatheta} \)P\(t,\lambda,I\(\frac{\gamma_{k+1}^-}{\lambda}\),I\(\frac{\gamma_k^+}{\lambda}\) \) \right.\nonumber\\
    &\quad\quad\quad  +\(1+\frac{a}{\sigmatheta}\)\(\int_t^T\(r_s-\frac{\|\btheta_s\|_2^2}{2} \)\,\d s P\(t,\lambda,I\(\frac{\gamma_{k+1}^-}{\lambda}\), I\(\frac{\gamma_k^+}{\lambda} \) \) \right.\nonumber\\
    &\quad\quad\quad\quad\left.\left.\left. -\sqrt{\int_t^T\|\btheta_s\|_2^2\,\d s} Q\(t,\lambda,I\(\frac{\gamma_{k+1}^-}{\lambda}\), I\(\frac{\gamma_k^+}{\lambda}\) \) \) \right\}\right\}\id_{\left\{R_{k}=\infty \right\} },
\end{align}
where $\tilde{k}:= k\exp\{(a+b)\Phi^{-1}(\bar{z})\}$. Finally, we define the following random variable, which is the optimal terminal wealth without the aspiration constraint:
\begin{equation*}
    \begin{aligned}
    Y^* 
    &= \sum_{k=0}^n a_k \id_{\left\{ \lambda^*\hat{\varphi}'\(1-w\(F_{\xi_{\boldsymbol{\theta}}}\(\xi_{\boldsymbol{\theta}}\) \)\) \in\(\gamma_k^+,\gamma_k^-\) \right\}} \\
    &\quad + \sum_{k=0}^n \left( A_k + \left(\frac{\gamma_k^{+}}{\lambda^*\hat{\varphi}'\(1-w\(F_{\xi_{\boldsymbol{\theta}}}\(\xi_{\boldsymbol{\theta}}\) \)\) }\right)^{\!R_k^{-1}}\!\!(a_k - A_k) \right) \id_{\left\{\lambda^*\hat{\varphi}'\(1-w\(F_{\xi_{\boldsymbol{\theta}}}\(\xi_{\boldsymbol{\theta}}\) \)\) \in \left(\gamma_{k+1}^{-}, \gamma_k^{+}\right)\right\}} \id_{\{ R_k \neq 0,\infty \}}\\
    &\quad + \sum_{k=0}^n \left(a_k + \frac{1}{\alpha_k} \log\left(\frac{\gamma_k^{+}}{\lambda^*\hat{\varphi}'\(1-w\(F_{\xi_{\boldsymbol{\theta}}}\(\xi_{\boldsymbol{\theta}}\) \)\)}\right)\right) \id_{\left\{\lambda^*\hat{\varphi}'\(1-w\(F_{\xi_{\boldsymbol{\theta}}}\(\xi_{\boldsymbol{\theta}}\) \)\) \in \left(\gamma_{k+1}^{-}, \gamma_k^{+}\right)\right\}} \id_{\left\{ R_k = \infty, A_k = -\infty, \alpha_k > 0 \right\}}.
\end{aligned}
\end{equation*}
Now we can discuss the four cases of the optimal terminal wealth based on the initial wealth level.

\noindent
(a) The initial wealth $x_0$ satisfies $x_0 + H(0) \notin \cup_{k=1}^n I_k $.
In this case, the optimal terminal wealth is given by
\begin{equation*}
    \begin{aligned}
        X_T^* = \[Y^* \vee L \]\id_{\{\xi_{\btheta} \leq F_{\xi_{\btheta}}^{-1}\(\alpha\) \}} + Y^* \id_{\{\xi_{\btheta}>F_{\xi_{\btheta}}^{-1}\(\alpha\) \}},
    \end{aligned}
\end{equation*}
where $\lambda^*$ is the unique Lagrange multiplier satisfying
\begin{equation*}
\int_0^1 Q_L(\lambda^*,y)\varphi'(y)\d y=x_0+H(0).
\end{equation*}
Then there are two cases: 

\noindent
(i) The Lagrange multiplier $\lambda^*$ satisfies $\lambda^* \varphi'(c) > \gamma_0^+$. The optimal wealth process is given by
\begin{align}
    X_t^* &=-H(t)+ \sum_{k=0}^n\( X_{t,k,\lambda^*}^D+X_{t,k,\lambda^*}^P+X_{t,k,\lambda^*}^E \)\nonumber\\
    &\quad +e^{-\int_t^T r_s\d s}L\[\Phi\(d\(t, I\(\frac{U'\(L\)}{\lambda^*}\) \) \) - \Phi\(d\(t,1-w\(\alpha\) \) \) \]\id_{\left\{1-w(\alpha)<I\(\frac{U'(L)}{\lambda^*} \) \right\} }.
\end{align}

\noindent
(ii) There exists $k^* \in \{0,\ldots, n-1\}$ such that $\gamma_{k^*}^+ > \lambda^* \varphi'\(c \)>\gamma_{k^*+1}^+ $. The optimal wealth process is given by
\begin{align}
    X_t^* &= -H(t)+\sum_{k=k^*+1}^n \(X_{t,k,\lambda^*}^D+X_{t,k,\lambda^*}^P+X_{t,k,\lambda^*}^E\)\nonumber\\
    &\quad+e^{-\int_t^T r_s\,\d s}L\[\Phi\(d\(t,I\(\frac{U'(L)}{\lambda^*}\)\)\)-\Phi\(d\(t,1-w(\alpha)\)\)\]\id_{\left\{1-w(\alpha)<I\left(\frac{U'(L)}{\lambda^*}\right)\right\}}\nonumber\\
    &\quad+e^{-\int_t^T r_s\,\d s}\Bigg\{\left(A_{k^*}+\left(\frac{\gamma_{k^*}^+}{\lambda^*\varphi'(c)}\right)^{R_{k^*}^{-1}}\(a_{k^*}-A_{k^*}\)\right)P\(t,\lambda^*,c,0\)\id_{\left\{c<I\left(\gamma_{k^*+1}^-/\lambda^*\right)\right\}}\nonumber\\
    &\quad\quad+A_{k^*}P\(t,\lambda^*,\max\left\{I\left(\frac{\gamma_{k^*+1}^-}{\lambda^*}\right),c\right\},c\)\nonumber\\
    &\quad\quad+\(a_{k^*}-A_{k^*}\)B_{t,k^*}\left(\frac{\tilde{k}\gamma_{k^*}^+}{\lambda^*}\right)^{R_{k^*}^{-1}}\xi_t^{-c_{k^*}}
    P_{k^*}\(t,\lambda^*,\max\left\{I\left(\frac{\gamma_{k^*+1}^-}{\lambda^*}\right),c\right\},c\)\Bigg\}\id_{\left\{R_{k^*}\neq0,\infty\right\}}\nonumber\\
    &\quad+e^{-\int_t^T r_s\,\d s}\Bigg\{\left(a_{k^*}+\frac{1}{\alpha_{k^*}}\log\left(\frac{\gamma_{k^*}^+}{\lambda^*\varphi'(c)}\right)\right)P\(t,\lambda^*,c,0\)
    \id_{\left\{c<I\left(\gamma_{k^*+1}^-/\lambda^*\right)\right\}}\nonumber\\
    &\quad\quad+a_{k^*}P\(t,\lambda^*,\max\left\{I\left(\frac{\gamma_{k^*+1}^-}{\lambda^*}\right),c\right\},c\)\nonumber\\
    &\quad\quad+\frac{1}{\alpha_{k^*}}\Bigg\{\left[\log\left(\frac{\tilde{k}\gamma_{k^*}^+}{\lambda^*\xi_t^{1+\frac{a}{\sigmatheta}}}\right)+\frac{a\mutheta}{\sigmatheta}
    +\left(1+\frac{a}{\sigmatheta}\right)\int_t^T\left(r_s-\frac{\|\btheta_s\|_2^2}{2}\right)\,\d s\right]\nonumber\\
    &\quad\quad\quad\times P\(t,\lambda^*,\max\left\{I\left(\frac{\gamma_{k^*+1}^-}{\lambda^*}\right),c\right\},c\)\nonumber\\
    &\quad\quad\quad-\left(1+\frac{a}{\sigmatheta}\right)\sqrt{\int_t^T\|\btheta_s\|_2^2\,\d s}\,
    Q\(t,\lambda^*,\max\left\{I\left(\frac{\gamma_{k^*+1}^-}{\lambda^*}\right),c\right\},c\)\Bigg\}\Bigg\}
    \id_{\left\{R_{k^*}=\infty\right\}} .
\end{align}

\noindent
(b) The initial wealth $x_0$ satisfies $x_0+ H(0) = x_{L,k_0}$ for some $k_0 \in \Lambda_n$.
The optimal terminal wealth is given by
\begin{equation*}
    \begin{aligned}
        X_T^* = \[Y^* \vee L \]\id_{\{\xi_{\btheta} \leq F_{\xi_{\btheta}}^{-1}\(\alpha\) \}} + Y^* \id_{\{\xi_{\btheta}>F_{\xi_{\btheta}}^{-1}\(\alpha\) \}}.
    \end{aligned}
\end{equation*}
Then the optimal wealth process is given by
\begin{align}\label{eq:lower X_t^*}
    X_t^* &=-H(t)+e^{-\int_t^T r_s \d s} a_{k_0} P(t,\lambda_{k_0},c,0)+ \sum_{k=k_0+1}^n\( X_{t,k,\lambda_{k_0}}^D+X_{t,k,\lambda_{k_0}}^P+X_{t,k,\lambda_{k_0}}^E \)\nonumber\\
    &\quad +e^{-\int_t^T r_s\d s}L\[\Phi\(d\(t, I\(\frac{U'\(L\)}{\lambda_{k_0}}\) \) \) - \Phi\(d\(t,1-w\(\alpha\) \) \) \]\id_{\left\{1-w(\alpha)<I\(\frac{U'(L)}{\lambda_{k_0}} \) \right\} }.
\end{align}

\noindent
(c) The initial wealth $x_0$ satisfies $x_0+ H(0) = x_{R,k_0}$ for some $k_0 \in \Lambda_n$.
The optimal terminal wealth is given by
\begin{equation*}
    \begin{aligned}
        X_T^* &= \[\(Y^*+\sum_{k=0}^{n-1}\(a_{k+1} - a_k\)\id_{\left\{\lambda_{k_0}\hat{\varphi}'\(1-w\(F_{\xi_{\boldsymbol{\theta}}}\(\xi_{\boldsymbol{\theta}}\)\)\)=\gamma_k^+=\gamma_{k+1}^- \right\}} \)\vee L \]\id_{\{\xi_{\btheta} \leq F_{\xi_{\btheta}}^{-1}\(\alpha\) \}}\\
        &\quad + \(Y^*+\sum_{k=0}^{n-1}\(a_{k+1} - a_k\)\id_{\left\{\lambda_{k_0}\hat{\varphi}'\(1-w\(F_{\xi_{\boldsymbol{\theta}}}\(\xi_{\boldsymbol{\theta}}\)\)\)=\gamma_k^+=\gamma_{k+1}^- \right\}} \) \id_{\{\xi_{\btheta} > F_{\xi_{\btheta}}^{-1}\(\alpha\) \}}.
    \end{aligned}
\end{equation*}
Then the optimal wealth process is given by
\begin{align}\label{eq:upper X_t^*}
    X_t^* &=-H(t)+ e^{-\int_t^T r_s \d s} a_{k_0+1}P(t,\lambda_{k_0},c,0)+ \sum_{k=k_0+1}^n\( X_{t,k,\lambda_{k_0}}^D+X_{t,k,\lambda_{k_0}}^P+X_{t,k,\lambda_{k_0}}^E \)\nonumber\\
    &\quad +e^{-\int_t^T r_s\d s}L\[\Phi\(d\(t, I\(\frac{U'\(L\)}{\lambda_{k_0}}\) \) \) - \Phi\(d\(t,1-w\(\alpha\) \) \) \]\id_{\left\{1-w(\alpha)<I\(\frac{U'(L)}{\lambda_{k_0}} \) \right\} }.
\end{align}

\noindent
(d) The initial wealth $x_0$ satisfies $x_0+ H(0) \in \cup_{k=1}^n I_k $. In this case, the present Lagrange construction does not identify an attainable optimizer.
\end{proposition}

Proposition \ref{prop:optimal wealth process} provides a closed-form representation of the optimal terminal payoff and the associated replicating wealth process. The random variable $Y^*$ represents the distorted quantile optimizer before the aspiration constraint is imposed. The aspiration constraint then modifies this unconstrained payoff by raising the terminal wealth to the level $L$ in the quantile region where the probability floor is binding. Hence, the optimal terminal wealth can be interpreted as the unconstrained distorted payoff adjusted by a benchmark-induced correction.

The proposition also clarifies the economic structure of the optimal wealth process. The terms $X^D_{t,k,\lambda}$, $X^P_{t,k,\lambda}$, and $X^E_{t,k,\lambda}$ correspond respectively to the linear, power, and exponential segments of the PHARA utility. These components reflect how the local shape of the effective utility determines the replicating value across different wealth regions. In addition, the aspiration constraint generates a correction term. This term appears only when the probability floor is active, and it represents the additional wealth required to ensure that the terminal payoff reaches the benchmark level $L$ in the relevant quantile region.

\begin{theorem}\label{thm:optimal portfolio}
Suppose that the concave envelope $U^{**}$ has the form in Definition \ref{def_PHARA} and $R_k\in[0,\infty]$ for each $k\in\{0,1,\ldots,n\}$. For fixed $t\in[0,T)$, $k=\{0,\ldots,n-1\}$ and $\lambda >0$, define the following processes: 
\begin{align}
\boldsymbol{\pi}_{t,k,\lambda}^D &:= -\(\boldsymbol{\sigma}_t\boldsymbol{\sigma}_t^\top\)^{-1}\(\boldsymbol{\mu}_t-r_t\mathbf{1}_m\)\frac{e^{-\int_t^T r_s \,\d s} }{\sqrt{\int_t^T\|\btheta_s\|_2^2\,\d s}} a_k Q\(t,\lambda,I\(\frac{\gamma_k^+}{\lambda} \), I\(\frac{\gamma_k^-}{\lambda}\) \),  \nonumber\\
\boldsymbol{\pi}_{t,k,\lambda}^P &:= -\(\boldsymbol{\sigma}_t\boldsymbol{\sigma}_t^\top\)^{-1}\(\boldsymbol{\mu}_t-r_t\mathbf{1}_m\)e^{-\int_t^T r_s \,\d s}\left\{\frac{A_k }{\sqrt{\int_t^T\|\btheta_s\|_2^2\,\d s}}Q\(t,\lambda,I\(\frac{\gamma_{k+1}^-}{\lambda} \), I\(\frac{\gamma_k^+}{\lambda} \) \) \right.\nonumber\\
&\quad\quad -\(a_k-A_k\)B_{t,k}\(\frac{\tilde{k}\gamma_k^+}{\lambda} \)^{R_k^{-1}}\xi_t^{-c_k}\Bigg\{ c_k P_k\(t,\lambda, I\(\frac{\gamma_{k+1}^-}{\lambda}\), I\(\frac{\gamma_k^+}{\lambda} \) \)\nonumber\\
&\quad\quad\quad \left.\left. -\frac{1}{\sqrt{\int_t^T\|\btheta_s\|_2^2\,\d s}}Q_k\(t,\lambda,I\(\frac{\gamma_{k+1}^-}{\lambda} \), I\(\frac{\gamma_k^+}{\lambda} \) \)\right\}\right\}\id_{\left\{R_k\neq0,\infty \right\} } ,\nonumber\\
\boldsymbol{\pi}_{t,k,\lambda}^E &:=-\(\boldsymbol{\sigma}_t\boldsymbol{\sigma}_t^\top\)^{-1}\(\boldsymbol{\mu}_t-r_t\mathbf{1}_m\)e^{-\int_t^T r_s\,\d s}\left\{\frac{a_k}{\sqrt{\int_t^T\|\btheta_s\|_2^2\,\d s}}Q\(t,\lambda,I\(\frac{\gamma_{k+1}^-}{\lambda}\),I\(\frac{\gamma_k^+}{\lambda}\)\)\right.\nonumber\\
&\quad+\frac{1}{\alpha_k}\left\{\left[\frac{1}{\sqrt{\int_t^T\|\btheta_s\|_2^2\,\d s}}\left(\log\left(\frac{\tilde{k}\gamma_k^+}{\lambda\xi_t^{1+\frac{a}{\sigmatheta}}}\right)+\frac{a\mutheta}{\sigmatheta}\right)+\left(1+\frac{a}{\sigmatheta}\right)\frac{\int_t^T\left(r_s-\|\btheta_s\|_2^2/2\right)\,\d s}{\sqrt{\int_t^T\|\btheta_s\|_2^2\,\d s}}\right]\right.\nonumber\\
&\quad\quad\times Q\(t,\lambda,I\(\frac{\gamma_{k+1}^-}{\lambda}\),I\(\frac{\gamma_k^+}{\lambda}\)\)-\left(1+\frac{a}{\sigmatheta}\right)P\(t,\lambda,I\(\frac{\gamma_{k+1}^-}{\lambda}\),I\(\frac{\gamma_k^+}{\lambda}\)\)\nonumber\\
&\quad\quad+\left(1+\frac{a}{\sigmatheta}\right)\left[\left(d\(t,I\(\frac{\gamma_{k+1}^-}{\lambda}\)\)\vee d\(t,I\(\frac{U'(L)}{\lambda}\)\)\right)\Phi'\left(d\(t,I\(\frac{\gamma_{k+1}^-}{\lambda}\)\)\vee d\(t,I\(\frac{U'(L)}{\lambda}\)\)\right)\right.\nonumber\\
&\quad\quad-\left(d\(t,I\(\frac{\gamma_k^+}{\lambda}\)\)\vee d\(t,I\(\frac{U'(L)}{\lambda}\)\)\right)\Phi'\left(d\(t,I\(\frac{\gamma_k^+}{\lambda}\)\)\vee d\(t,I\(\frac{U'(L)}{\lambda}\)\)\right)\nonumber\\
&\quad\quad+\left(d\(t,I\(\frac{\gamma_{k+1}^-}{\lambda}\)\)\wedge d\(t,1-w(\alpha)\)\right)\Phi'\left(d\(t,I\(\frac{\gamma_{k+1}^-}{\lambda}\)\)\wedge d\(t,1-w(\alpha)\)\right)\nonumber\\
&\quad\quad\left.\left.\left.-\left(d\(t,I\(\frac{\gamma_k^+}{\lambda}\)\)\wedge d\(t,1-w(\alpha)\)\right)\Phi'\left(d\(t,I\(\frac{\gamma_k^+}{\lambda}\)\)\wedge d\(t,1-w(\alpha)\)\right)\right]\right\}\right\}\id_{\left\{R_k=\infty\right\}} .
\end{align}

There are four cases of the optimal portfolio process depending on the initial wealth $x_0$.

\noindent
(a) The initial wealth $x_0$ satisfies $x_0+ H(0) \notin \cup_{k=1}^n I_k $. Then we consider two scenarios.

\noindent
(i) The Lagrange multiplier $\lambda^*$ satisfies $\lambda^* \varphi'(c) > \gamma_0^+$. The optimal portfolio process is given by
\begin{align}
    \boldsymbol{\pi}_t^* &=\sum_{k=0}^n \(\boldsymbol{\pi}_{t,k,\lambda^*}^D+\boldsymbol{\pi}_{t,k,\lambda^*}^P+\boldsymbol{\pi}_{t,k,\lambda^*}^E \)-\(\boldsymbol{\sigma}_t\boldsymbol{\sigma}_t^\top\)^{-1}\(\boldsymbol{\mu}_t-r_t\mathbf{1}_m\)\frac{e^{-\int_t^T r_s \,\d s} L}{\sqrt{\int_t^T\|\btheta_s\|_2^2\,\d s}}\nonumber\\
    &\quad  \times \[\Phi'\( d\(t, I\(\frac{U'(L)}{\lambda^*} \) \) \)-\Phi'\( d\(t, 1-w(\alpha)\) \) \]\id_{\left\{1-w(\alpha)<I\(\frac{U'(L)}{\lambda^*} \) \right\} }.
\end{align}

\noindent
(ii) There exists $k^* \in \{0,\ldots, n-1\}$ such that $\gamma_{k^*}^+> \lambda^* \(\varphi \)'\(c \)>\gamma_{k^*+1}^+ $. The optimal portfolio process is given by
\begin{align}
    \boldsymbol{\pi}_t^* &= \sum_{k=k^*+1}^n \(\boldsymbol{\pi}_{t,k,\lambda^*}^D+\boldsymbol{\pi}_{t,k,\lambda^*}^P+\boldsymbol{\pi}_{t,k,\lambda^*}^E \)-\(\boldsymbol{\sigma}_t\boldsymbol{\sigma}_t^\top\)^{-1}\(\boldsymbol{\mu}_t-r_t\mathbf{1}_m\)\nonumber\\
    &\quad\times \frac{e^{-\int_t^T r_s \,\d s} L}{\sqrt{\int_t^T\|\btheta_s\|_2^2\,\d s}}  \[\Phi'\( d\(t, I\(\frac{U'(L)}{\lambda^*} \) \) \)-\Phi'\(d\(t, 1-w(\alpha)\) \) \]\id_{\left\{1-w(\alpha)<I\(\frac{U'(L)}{\lambda^*} \) \right\} }\nonumber\\
    &\quad-\(\boldsymbol{\sigma}_t\boldsymbol{\sigma}_t^\top\)^{-1}\(\boldsymbol{\mu}_t-r_t\mathbf{1}_m\)e^{-\int_t^T r_s \,\d s}\left\{\left\{\frac{1}{\sqrt{\int_t^T\|\btheta_s\|_2^2\,\d s}} \(A_{k^{*}}+ \(\frac{\gamma_{k^{*}}^+ }{\lambda^* \varphi'(c)}\)^{R_{k^{*}}^{-1}}\(a_{k^{*}}-A_{k^{*}} \) \) \right.\right.\nonumber\\
    &\quad\times Q\(t,\lambda^*,c,0  \)\id_{\left\{c < I\(\gamma_{k^*+1}^-/\lambda^* \) \right\}} +\frac{A_{k^*} }{\sqrt{\int_t^T\|\btheta_s\|_2^2\,\d s}}Q\(t,\lambda^*,\max\left\{ I\(\frac{\gamma_{k^*+1}^-}{\lambda^*} \),c\right\}, c \) \nonumber\\
    &\quad -\(a_{k^*}-A_{k^*}\)B_{t,{k^*}}\(\frac{\tilde{k}\gamma_{k^*}^+}{\lambda^*} \)^{R_{k^*}^{-1}}\xi_t^{-c_{k^*}}\left\{ c_{k^*} P_{k^*}\(t,\lambda^*,\max\left\{I\(\frac{\gamma_{k^*+1}^-}{\lambda^*}\),c\right\}, c \)\right.\nonumber\\
    &\quad\quad \left.\left. -\frac{1}{\sqrt{\int_t^T\|\btheta_s\|_2^2\,\d s}}Q_{k^*}\(t,\lambda^*,\max\left\{I\(\frac{\gamma_{k^*+1}^-}{\lambda^*} \),c\right\}, c \)\right\}\right\}\id_{\left\{R_{k^*} \neq 0, \infty \right\}} \nonumber\\
    &\quad+\Bigg\{\frac{1}{\sqrt{\int_t^T\|\btheta_s\|_2^2\,\d s}}\left(a_{k^*}+\frac{1}{\alpha_{k^*}}\log\left(\frac{\gamma_{k^*}^+}{\lambda^*\varphi'(c)}\right)\right)Q(t,\lambda^*,c,0)\id_{\left\{c<I\left(\gamma_{k^*+1}^-/\lambda^*\right)\right\}}\nonumber\\
    &\quad+\frac{a_{k^*}}{\sqrt{\int_t^T\|\btheta_s\|_2^2\,\d s}}Q\left(t,\lambda^*,\max\left\{I\left(\frac{\gamma_{k^*+1}^-}{\lambda^*}\right),c\right\},c\right)+\frac{1}{\alpha_{k^*}}\Bigg\{\Bigg[\frac{1}{\sqrt{\int_t^T\|\btheta_s\|_2^2\,\d s}}\left(\log\left(\frac{\tilde{k}\gamma_{k^*}^+}{\lambda^*\xi_t^{1+\frac{a}{\sigmatheta}}}\right)+\frac{a\mutheta}{\sigmatheta}\right)\nonumber\\
    &\quad\quad+\left(1+\frac{a}{\sigmatheta}\right)\frac{\int_t^T\left(r_s-\|\btheta_s\|_2^2/2\right)\,\d s}{\sqrt{\int_t^T\|\btheta_s\|_2^2\,\d s}}\Bigg]Q\left(t,\lambda^*,\max\left\{I\left(\frac{\gamma_{k^*+1}^-}{\lambda^*}\right),c\right\},c\right)\nonumber\\
    &\quad\quad-\left(1+\frac{a}{\sigmatheta}\right)P\left(t,\lambda^*,\max\left\{I\left(\frac{\gamma_{k^*+1}^-}{\lambda^*}\right),c\right\},c\right)\nonumber\\
    &\quad\quad+\left(1+\frac{a}{\sigmatheta}\right)\Bigg[\left(d\left(t,\max\left\{I\left(\frac{\gamma_{k^*+1}^-}{\lambda^*}\right),c\right\}\right)\vee d\left(t,I\left(\frac{U'(L)}{\lambda^*}\right)\right)\vee d(t,1-w(\alpha))\right)\nonumber\\
    &\quad\quad\times\Phi'\left(d\left(t,\max\left\{I\left(\frac{\gamma_{k^*+1}^-}{\lambda^*}\right),c\right\}\right)\vee d\left(t,I\left(\frac{U'(L)}{\lambda^*}\right)\right)\vee d(t,1-w(\alpha))\right)\nonumber\\
    &\quad\quad-\left(d(t,c)\vee d\left(t,I\left(\frac{U'(L)}{\lambda^*}\right)\right)\vee d(t,1-w(\alpha))\right)\Phi'\left(d(t,c)\vee d\left(t,I\left(\frac{U'(L)}{\lambda^*}\right)\right)\vee d(t,1-w(\alpha))\right)\nonumber\\
    &\quad\quad+\left(d\left(t,\max\left\{I\left(\frac{\gamma_{k^*+1}^-}{\lambda^*}\right),c\right\}\right)\wedge d(t,1-w(\alpha))\right)\Phi'\left(d\left(t,\max\left\{I\left(\frac{\gamma_{k^*+1}^-}{\lambda^*}\right),c\right\}\right)\wedge d(t,1-w(\alpha))\right)\nonumber\\
    &\quad\quad-\left(d(t,c)\wedge d(t,1-w(\alpha))\right)\Phi'\left(d(t,c)\wedge d(t,1-w(\alpha))\right)\Bigg]\Bigg\}\Bigg\}\id_{\left\{R_{k^*}=\infty\right\}} .
\end{align}

\noindent
(b) The initial wealth $x_0$ satisfies $x_0+ H(0)=x_{L,k_0} $ for some $k_0 \in \Lambda_n$. The optimal portfolio process is given by
\begin{align}
    \boldsymbol{\pi}_t^* &=-\(\boldsymbol{\sigma}_t\boldsymbol{\sigma}_t^\top\)^{-1}\(\boldsymbol{\mu}_t-r_t\mathbf{1}_m\)\frac{ e^{-\int_t^T r_s\,\d s}a_{k_0}}{\sqrt{\int_t^T\|\btheta_s\|_2^2\,\d s}} Q(t,\lambda_{k_0},c,0)\nonumber\\
    &\quad+\sum_{k=k_0+1}^n \(\boldsymbol{\pi}_{t,k,\lambda_{k_0}}^D+\boldsymbol{\pi}_{t,k,\lambda_{k_0}}^P+\boldsymbol{\pi}_{t,k,\lambda_{k_0}}^E \)-\(\boldsymbol{\sigma}_t\boldsymbol{\sigma}_t^\top\)^{-1}\(\boldsymbol{\mu}_t-r_t\mathbf{1}_m\)\nonumber\\
    &\quad\times \frac{e^{-\int_t^T r_s \,\d s} L}{\sqrt{\int_t^T\|\btheta_s\|_2^2\,\d s}}  \[\Phi'\(d\(t, I\(\frac{U'(L)}{\lambda_{k_0}} \) \) \)-\Phi'\(d\(t, 1-w(\alpha)\) \) \]\id_{\left\{1-w(\alpha)<I\(\frac{U'(L)}{\lambda_{k_0}} \) \right\} }.
\end{align}

\noindent
(c) The initial wealth $x_0$ satisfies $x_0+ H(0)=x_{R,k_0} $ for some $k_0 \in \Lambda_n$. The optimal portfolio process is given by
\begin{align}
    \boldsymbol{\pi}_t^* &=-\(\boldsymbol{\sigma}_t\boldsymbol{\sigma}_t^\top\)^{-1}\(\boldsymbol{\mu}_t-r_t\mathbf{1}_m\)\frac{ e^{-\int_t^T r_s\,\d s}a_{k_0+1}}{\sqrt{\int_t^T\|\btheta_s\|_2^2\,\d s}} Q(t,\lambda_{k_0},c,0)\nonumber\\
    &\quad+\sum_{k=k_0+1}^n \(\boldsymbol{\pi}_{t,k,\lambda_{k_0}}^D+\boldsymbol{\pi}_{t,k,\lambda_{k_0}}^P+\boldsymbol{\pi}_{t,k,\lambda_{k_0}}^E \)-\(\boldsymbol{\sigma}_t\boldsymbol{\sigma}_t^\top\)^{-1}\(\boldsymbol{\mu}_t-r_t\mathbf{1}_m\)\nonumber\\
    &\quad\times \frac{e^{-\int_t^T r_s \,\d s} L}{\sqrt{\int_t^T\|\btheta_s\|_2^2\,\d s}}  \[\Phi'\(d\(t, I\(\frac{U'(L)}{\lambda_{k_0}} \) \) \)-\Phi'\(d\(t, 1-w(\alpha)\) \) \]\id_{\left\{1-w(\alpha)<I\(\frac{U'(L)}{\lambda_{k_0}} \) \right\} }.
\end{align}

\noindent
(d) The initial wealth $x_0$ satisfies $x_0+ H(0)\in \cup_{k=1}^n I_k$. In this case, our method fails to provide an optimal solution.

\end{theorem}
Theorem \ref{thm:optimal portfolio} gives the optimal portfolio process corresponding to the wealth process in Proposition \ref{prop:optimal wealth process}. After the labor-capital adjustment, the wealth process is self-financing, so the portfolio can be obtained from the martingale representation. Equivalently, it follows by differentiating the replicating value with respect to the pricing kernel. Like the components in Proposition \ref{prop:optimal wealth process}, the three terms $\pi^D_{t,k,\lambda}$, $\pi^P_{t,k,\lambda}$, and $\pi^E_{t,k,\lambda}$ come from the linear, power, and exponential parts of the PHARA utility, respectively. Thus, the risky investment depends on both the market price of risk and the marginal shape of the effective utility induced by the participating contract.

The terms involving $L$ come from the aspiration constraint. They disappear when the probability floor is not binding. When the floor binds, these terms modify the risky position so that the terminal wealth can satisfy the benchmark requirement in the relevant quantile region. In this way, the formula separates the effects of the PHARA envelope, probability distortion, and the aspiration constraint on the optimal investment policy.

\section{Economic meanings and asymptotic analysis}\label{sec:motivation revisit}

\begin{figure}[h!]
    \centering
    \includegraphics[width=0.31\linewidth]{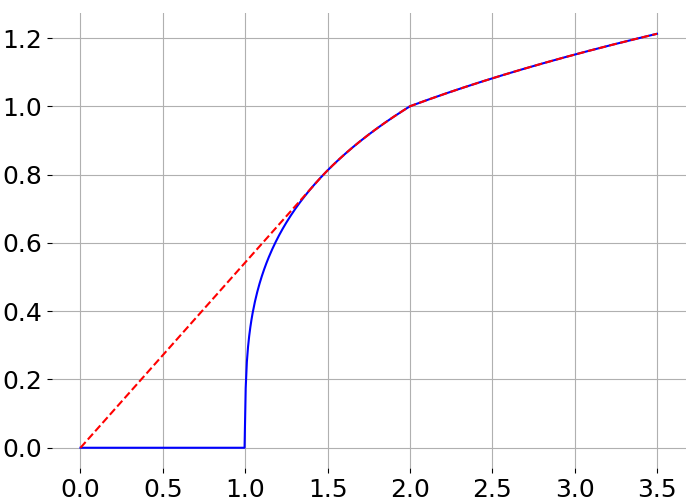}
    \includegraphics[width=0.31\linewidth]{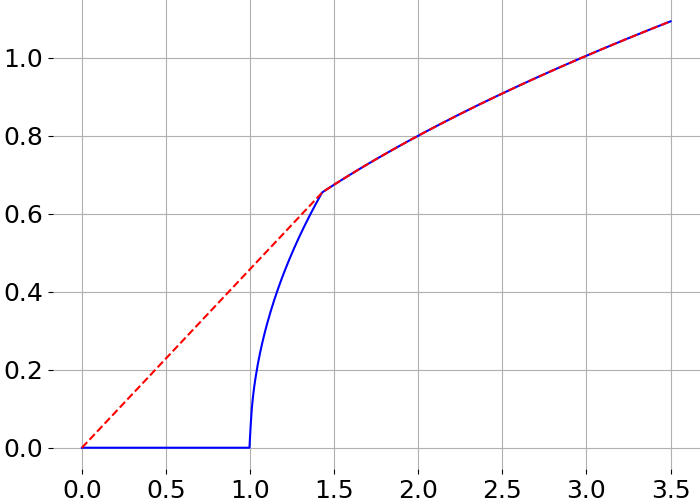}
    \includegraphics[width=0.31\linewidth]{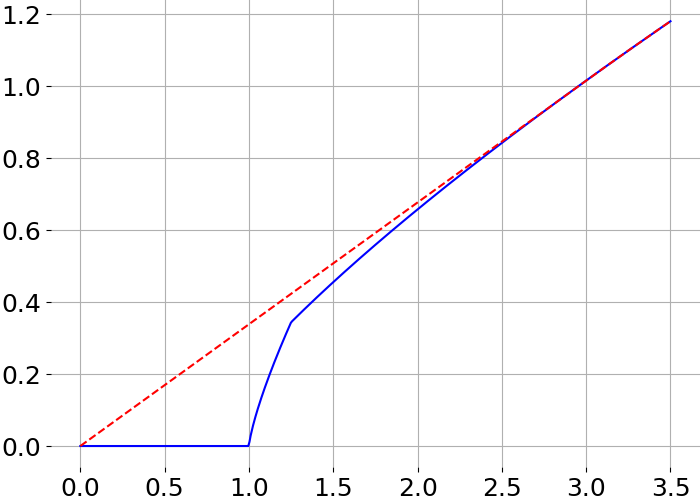}
    \caption{Three cases corresponding to $c_U < L_T^g/\eta$, $c_U = L_T^g/\eta$, and $c_U > L_T^g/\eta$, respectively.}
    \label{fig:three envelopes}
\end{figure}

In this section, we revisit the motivating example in Section \ref{sec:motivation}. We consider the third scenario where $B=0$. In this case, there are three forms of the concave envelopes based on the value of $L_T^g/\eta$ and $c_U$ defined as follows: 
\begin{equation}
    c_U := \sup_{x \geq 0}\{U^{**}(x) > U(x) \}.
\end{equation}
The three types of concave envelopes are shown in Figure \ref{fig:three envelopes}.

We take the second case for example. The concave envelope belongs to the PHARA family with $a_0 = 0$, $a_1 = c_U = L_T^g/\eta$, $a_2= \infty$. 
In this case, the optimal wealth process is given by 
\begin{equation*}
\begin{aligned}
    X_t^* &= e^{-\int_t^Tr_s \d s}\left\{a_1 P\(t, \lambda, I\(\frac{\gamma^+}{\lambda}\),I\(\frac{\gamma^-}{\lambda} \) \) + \(\frac{\(1-\delta\)L_T^g}{1-\delta\eta}P\(t,\lambda,1,I\(\frac{\gamma^+}{\lambda}\) \)\)\right.\\
    &\quad+\left.\(a_1 -\frac{\(1-\delta\)L_T^g}{1-\delta\eta} \)B_{t,1}\(\frac{\tilde{k}\gamma^+}{\lambda} \)^{\frac{1}{1-p}}\xi_t^{-\frac{a+\sigmatheta}{(1-p)\sigmatheta }}P_1\(t,\lambda, 1,I\(\frac{\gamma^+}{\lambda}\) \)\right\},
\end{aligned}
\end{equation*}
where $\gamma^+ := U'\(a_1+\) $ and $\gamma^-:= U(a_1)/a_1$.
The optimal portfolio is
\begin{equation*}
    \begin{aligned}
    \bpi_t^* &=-\(\boldsymbol{\sigma}_t\boldsymbol{\sigma}_t^\top\)^{-1}\(\boldsymbol{\mu}_t-r_t\mathbf{1}_m\) e^{-\int_t^Tr_s \d s}\left\{\frac{a_1}{\sqrt{\int_t^T\|\btheta_s\|_2^2\d s }} Q\(t, \lambda, I\(\frac{\gamma^+}{\lambda}\),I\(\frac{\gamma^-}{\lambda} \) \)\right.\\
    &\quad+\frac{\(1-\delta\)L_T^g}{\(1-\delta\eta\)\sqrt{\int_t^T\|\btheta_s\|_2^2\d s}}Q\(t,\lambda,1,I\(\frac{\gamma^+}{\lambda}\) \)-\(a_1 -\frac{\(1-\delta\)L_T^g}{1-\delta\eta} \)B_{t,1}\(\frac{\tilde{k}\gamma^+}{\lambda} \)^{\frac{1}{1-p}}\xi_t^{-\frac{a+\sigmatheta}{(1-p)\sigmatheta }}\\
    &\quad \left.\times\(\frac{a+\sigmatheta}{(1-p)\sigmatheta }P_1\(t,\lambda,1,I\(\frac{\gamma^+}{\lambda}\) \)-\frac{1}{\sqrt{\int_t^T \|\btheta_s\|_2^2\d s}}Q_1\(t,\lambda, 1,I\(\frac{\gamma^+}{\lambda}\)\) \)\right\}.
    \end{aligned}
\end{equation*}
Using the asymptotic analysis approach of  \cite{LL2024}, we obtain the asymptotic behaviors of the optimal results.
\begin{proposition}\label{prop:asymptotic}
As $\xi_t\rightarrow 0$, we have
\begin{equation*}
    X_t^*\rightarrow\infty,\quad\bpi_t^*\rightarrow \mathbf{\infty},\quad \frac{\bpi_t^*}{X_t^*}\rightarrow \(\boldsymbol{\sigma}_t\boldsymbol{\sigma}_t^\top\)^{-1}\(\boldsymbol{\mu}_t -r_t \mathbf{1}_m\) \cdot\frac{\sigmatheta+a}{(1-p)\sigmatheta}.
\end{equation*}
As $\xi_t \rightarrow \infty$, there exist two cases.\\
Case 1. $\lambda\varphi'(c)\leq\gamma^+$: In this case, we have
\begin{equation*}
    X_t^* \rightarrow \frac{\(1-\delta\)L_T^g}{1-\delta\eta},\quad \bpi_t^*\rightarrow \mathbf{0}, \quad \frac{\bpi_t^*}{X_t^*}\rightarrow \mathbf{0}. 
\end{equation*}
Case 2. $\lambda\varphi'(c)>\gamma^+$: In this case, we have
\begin{equation*}
    X_t^* \rightarrow 0, \quad \bpi_t^* \rightarrow \mathbf{0},\quad \frac{\bpi_t^*}{X_t^*}\rightarrow \boldsymbol{\infty}.
\end{equation*}

\end{proposition}

\begin{figure}[h!]
    \centering
    \includegraphics[width=0.7\linewidth]{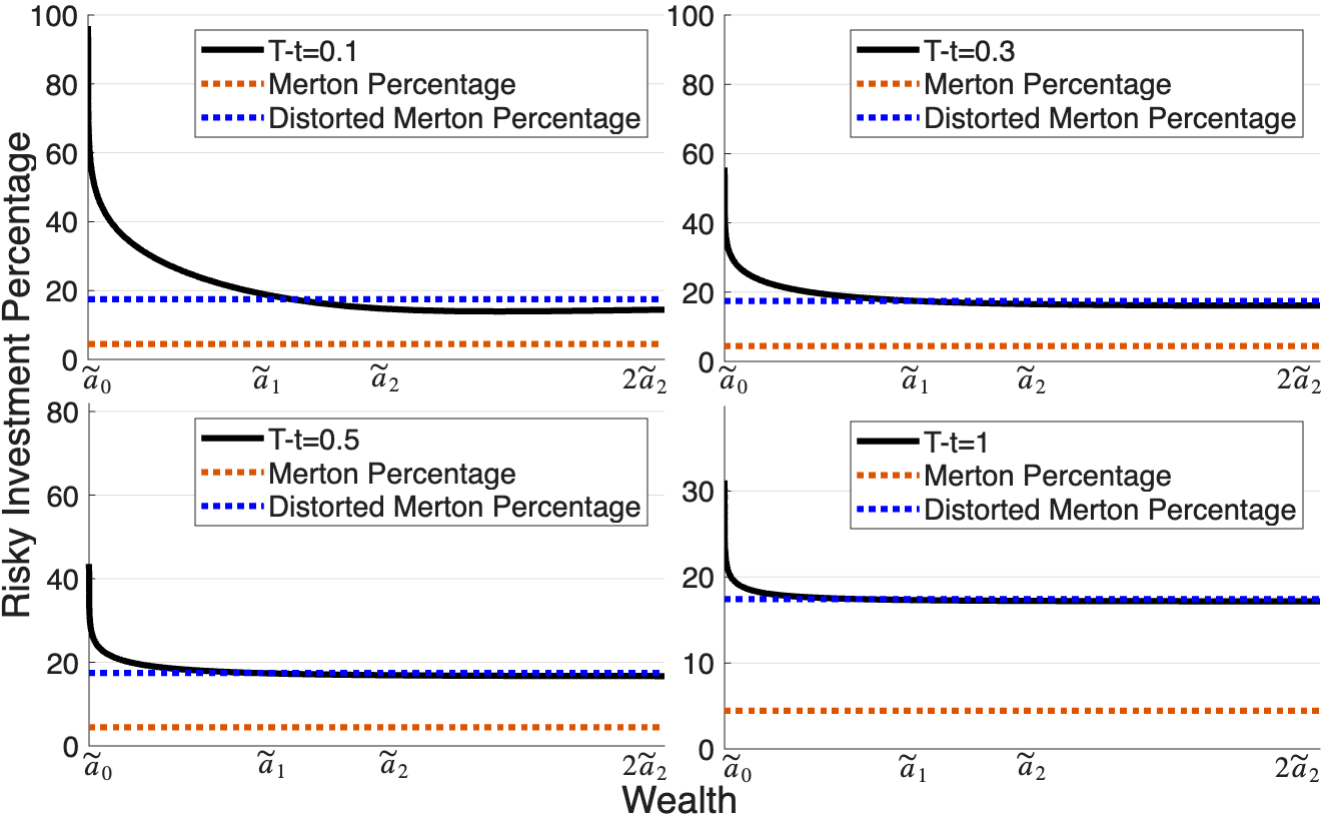}\\
    \includegraphics[width=0.7\linewidth]{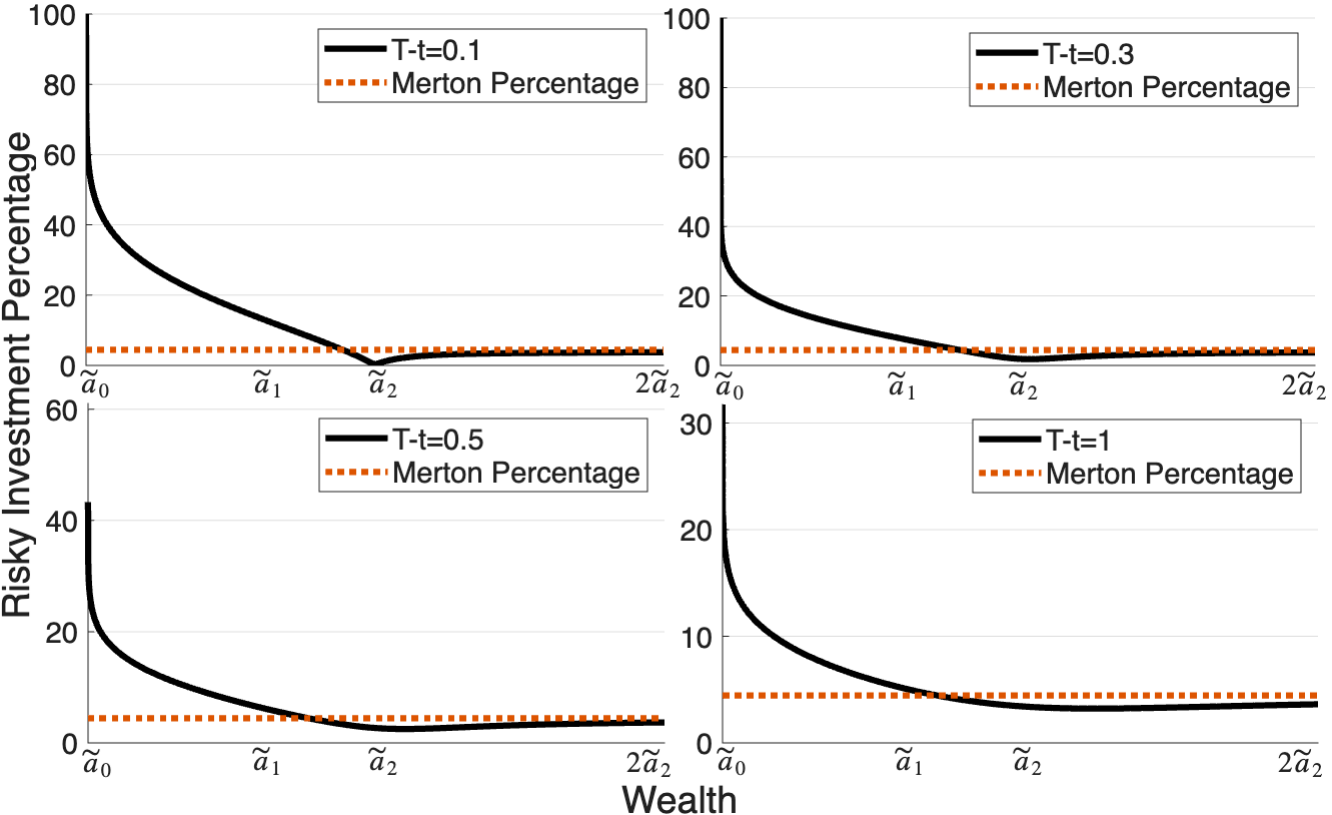}
    \caption{The parameters for the upper figure are given by $\mu = 0.07$, $r = 0.01$, $\sigma = 0.15$, $T = 1$, $a = 1.17$, $b = 0.86$, $\bar{z} = 0.33$, $L_T^g = 1$, $p = 0.4$, $\eta = 0.6$, $x_0 = 1$, while $a$, $b$, $\bar{z}$ are set to be $0$ for the lower figure. The Merton percentage is $\frac{\theta}{\sigma (1-p)}$ in both cases.}
    \label{fig:optimal pis xichange}
\end{figure}

This proposition will be shown explicitly in the following numerical examples.
The optimal investment strategies with and without distortion at each time points are shown in Figure \ref{fig:optimal pis xichange}.

We observe that in the non-distorted case, all optimal investment ratios converge to the Merton percentage, regardless of the time point. Under probability distortion, however, this convergence is disrupted: the investment ratios no longer converge to the Merton benchmark, but a distorted Merton line. A direct asymptotic analysis shows that as $\xi_t \rightarrow 0$, the investment ratio converges to $\frac{\theta}{\sigma}\cdot\frac{a + \sigma_{\theta}}{(1-p)\sigma_{\btheta}}$. Since $a, \sigma_\theta > 0$, the limit is always larger than the Merton ratio, indicating that the inverse S-shaped probability distortion consistently increases the insurers' risk exposure. Under an inverse S-shaped probability distortion, the insurer systematically overestimates the likelihood of small-probability events while underestimating the likelihood of high-probability ones. Since the risky asset carries a positive expected return, its value is more likely to rise than to fall. An insurer subject to probability distortion therefore overestimates the likelihood of an upward move and, as a result, becomes less inclined to engage in the lock-in behavior. This leads to a substantial increase in risky investment once the fund value exceeds the reference point.

\begin{figure}[h!]
    \centering
    \includegraphics[width=0.7\linewidth]{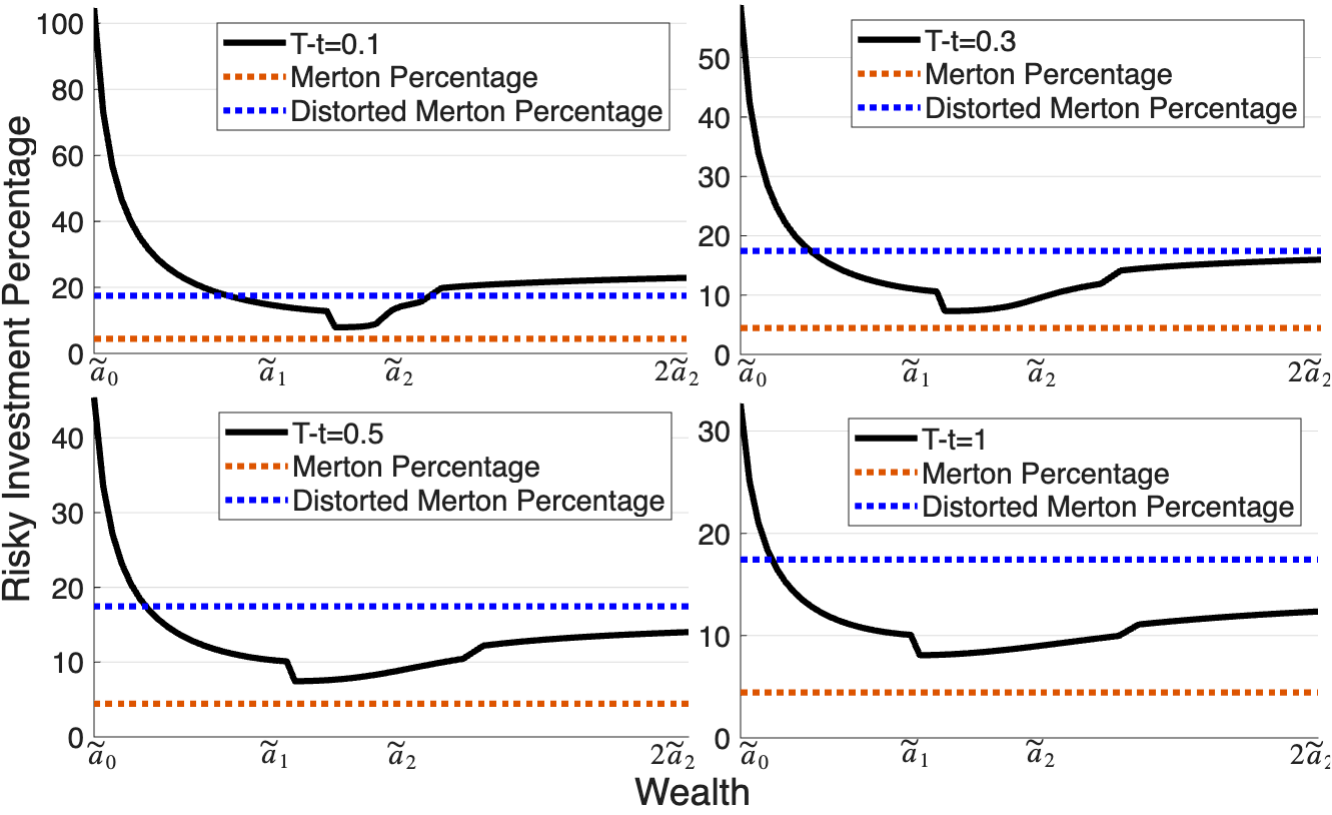}\\
    \includegraphics[width=0.7\linewidth]{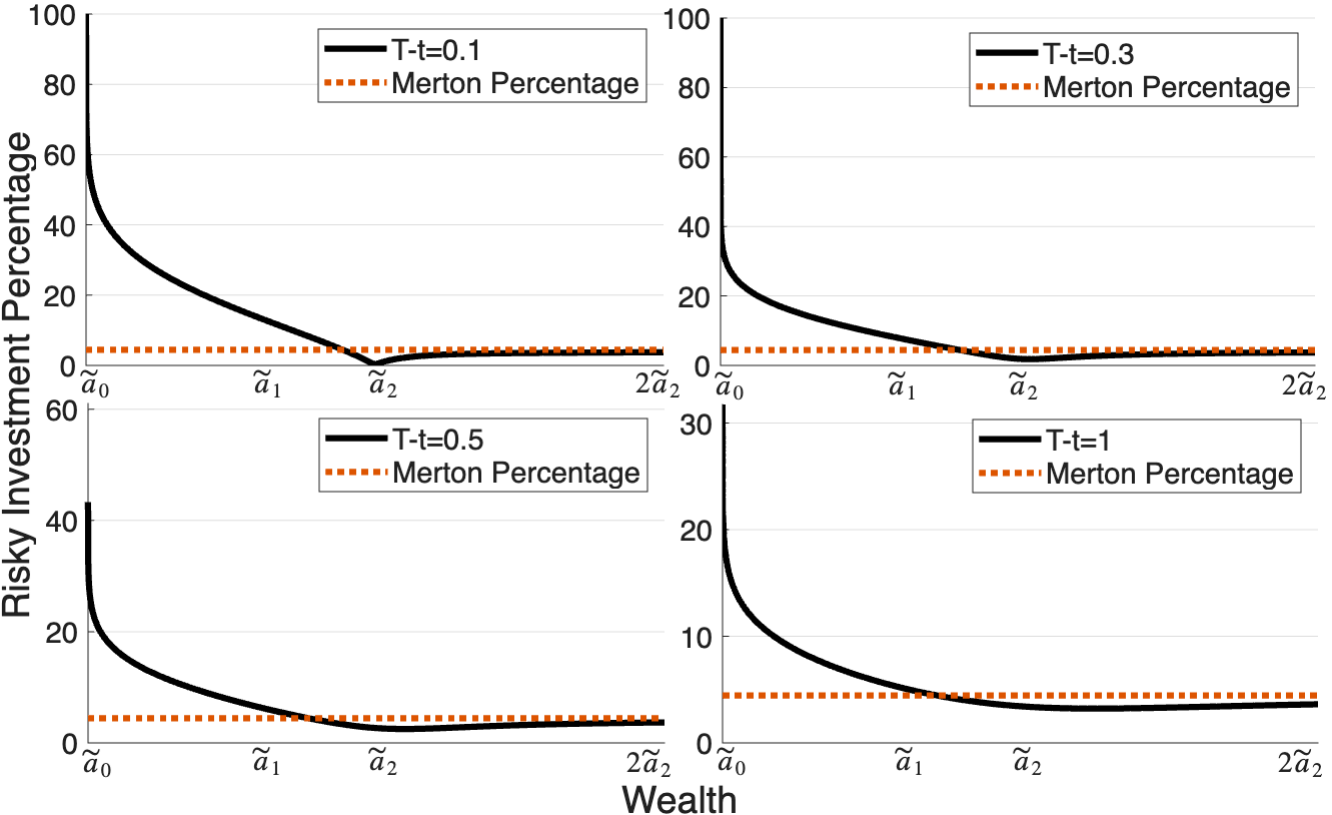}
    \caption{Optimal investment strategies with and without probability distortion starting at each time points. The parameters for the upper figure are given by $\mu = 0.07$, $r = 0.01$, $\sigma = 0.15$, $T = 1$, $a = 1.17$, $b = 0.86$, $\bar{z} = 0.33$, $L_T^g = 1$, $p = 0.4$, $\eta = 0.6$, while $a$, $b$, $\bar{z}$ are set to be $0$ for the lower figure. The Merton percentage is $\frac{\theta}{\sigma (1-p)}$ in both cases.}
    \label{fig:optimal pis nuchange}
\end{figure}

Moreover, the investment curves under distortion are less smooth than those in the non-distorted case. In particular, they exhibit sharp jumps around the non-differentiable point $\tilde{a}_2$. As shown in \cite{LLMV2024}, the risky investment percentage forms a valley at such non-differentiable points. In an actuarial context, the non-differentiable point can be interpreted as a reserve threshold. Insurers often adjust risk-taking when the wealth level approaches such regulatory or internal reference levels: they take more risk when the wealth is low to avoid breaching capital requirements, and switch to a lock-in strategy once the wealth exceeds the threshold. Probability distortion weakens this lock-in behavior. This fits with the increased risk-exposure attitude. The insurer is overly optimistic about the portfolio such that he/she overlooks the drop-off risk that may induce great loss from the incentive plan.
We also show the variation of the investment pattern for different investment horizons. In Figure \ref{fig:optimal pis nuchange}, we present the initial investment percentage for different investment horizons.

To distinguish Figures \ref{fig:optimal pis xichange}-\ref{fig:optimal pis nuchange}, we first introduce the undistorted problem
\begin{equation}\label{eq:numerical prob}
    \sup_{\bpi \in \mathcal{V}[0,T]} \E\[U\(X_T\) \],\quad X_0 = 1.
\end{equation}
For the distorted case, we keep the same market setting and initial wealth, but replace the expected-utility objective in \eqref{eq:numerical prob} by the distorted objective in Problem \eqref{eq: original}. Figure \ref{fig:optimal pis xichange} displays the optimal investment strategies under the distorted and undistorted objectives for $t=0,0.5,0.7,0.9$, with the terminal time fixed at $T=1$. Figure \ref{fig:optimal pis nuchange} displays the corresponding strategies at $t=0$ for $T=0.1,0.3,0.5,1$.

The most significant distinction between the distorted and undistorted investment strategies is the time-inconsistency. As illustrated in Figures \ref{fig:optimal pis xichange}–\ref{fig:optimal pis nuchange}, the undistorted investment curves coincide for different starting times. In this case, the investor's decision depends only on the remaining time to maturity, rather than on the calendar time itself. This property reflects time consistency and is consistent with the Bellman optimality principle, under which the dynamically optimal strategy remains optimal for any truncated subproblem.

However, when probability distortion is introduced, the Bellman principle no longer holds. The distorted objective functional induces dynamically inconsistent preferences, so that an investment strategy that is optimal at the initial time fails to remain optimal when the problem is reconsidered at a later date with the same remaining horizon. Consequently, the optimal investment policy becomes explicitly time-dependent, and the investment curves corresponding to different starting times no longer overlap. More precisely, Figure \ref{fig:optimal pis nuchange} compares optimization problems initiated at different calendar times while keeping the remaining horizon fixed. Under the classical expected utility criterion, these problems admit the same optimal strategy because only the time-to-maturity matters. In contrast, under probability distortion, starting the optimization at different dates leads to different optimal policies, providing a direct numerical illustration of time inconsistency. From an actuarial perspective, this loss of time consistency implies that the investor faces a discrepancy between ex ante optimality and interim optimality, highlighting the necessity of distinguishing between pre-committed strategies and closed-loop strategies in the presence of probability distortion.

\begin{figure}[h!]
    \centering
    \includegraphics[width=0.7\linewidth]{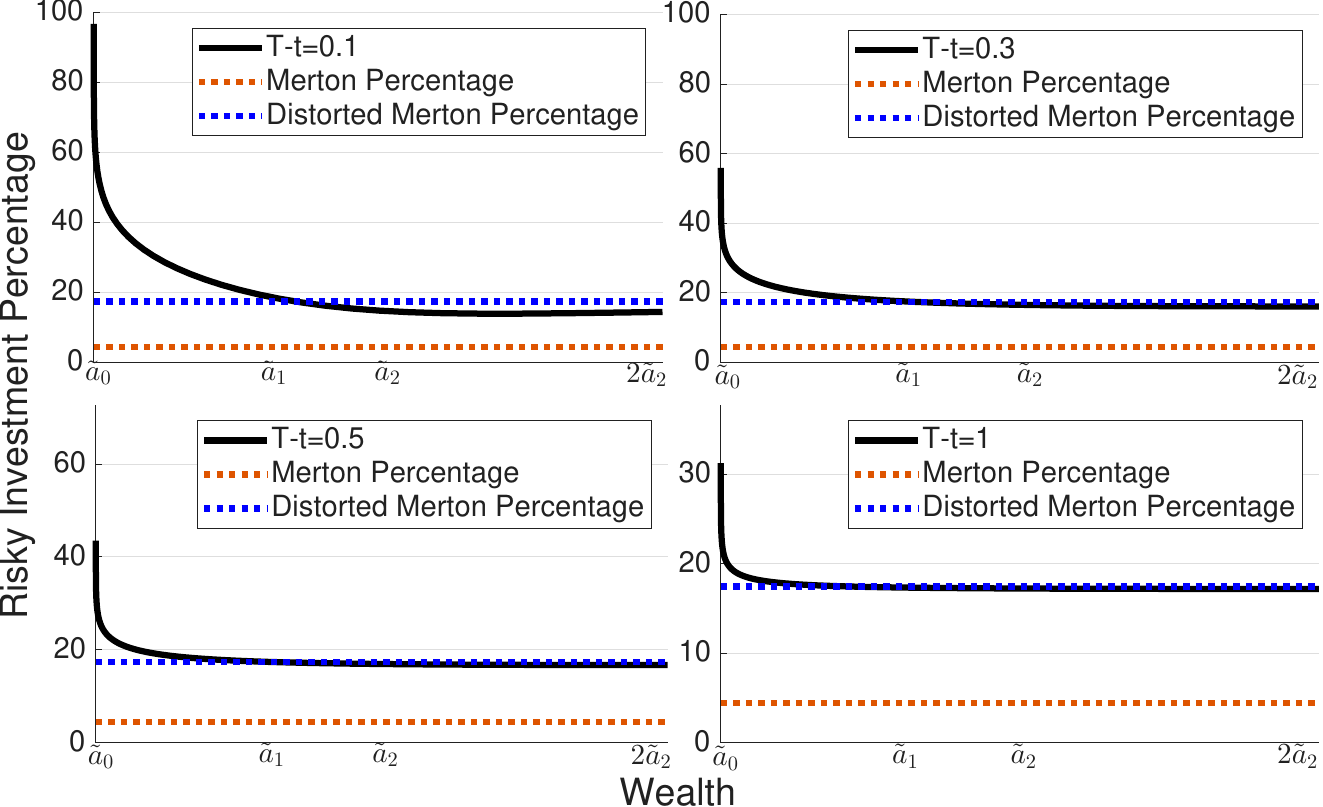}\\
    \includegraphics[width=0.7\linewidth]{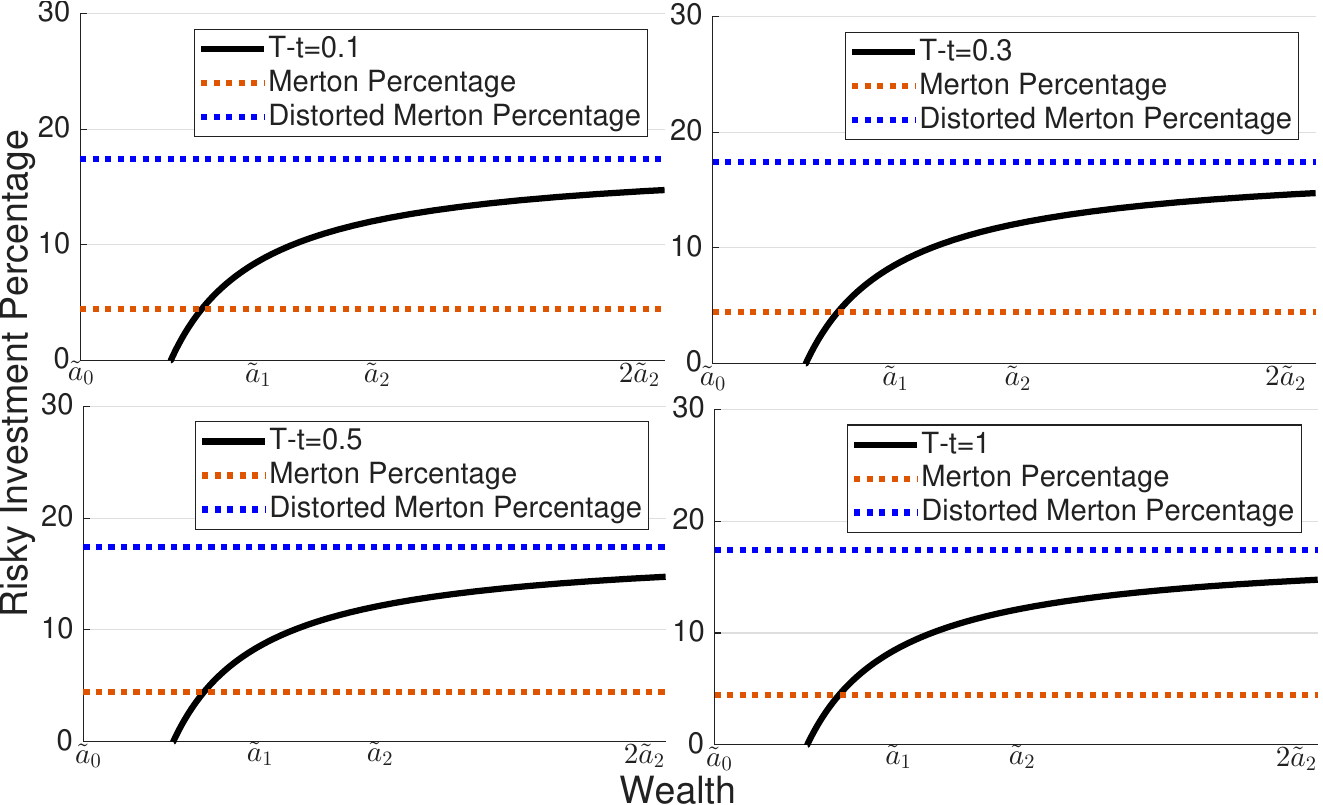}
    \caption{Optimal investment strategy sample paths under probability distortion starting at $t = 0$, $x_0 = 0.5, 3$, respectively. The parameters are the same as those in Figure \ref{fig:optimal pis nuchange}.}
    \label{fig: Bi figures}
\end{figure}
Finally, Figure \ref{fig: Bi figures} shows that the initial value can significantly shift the investment patterns. A notable difference between the investment patterns lies in the low-wealth investment percentage. With a low initial value, the insurer increases the allocation to risky assets to escape the low-wealth region. Provided a higher initial value, however, the insurer behaves in an excessively risk-averse manner, allocating almost entirely to the risk-free asset. Meanwhile, the investment strategy guarantees that the fund value will not drop below a positive level. This qualitative change in investment behavior is essentially driven by the emergence of a critical Lagrange multiplier associated with the probability distortion, as analyzed in \cite{Bi2023}. From an actuarial perspective, this mechanism highlights how initial capitalization and regulatory or solvency constraints can jointly lead to sharp regime switches in optimal investment policies, even when the underlying market dynamics remain unchanged.



\section{Concluding remarks}\label{sec:conclusion}

This paper establishes a unified quantile-based framework for optimal investment in participating insurance contracts under probability distortion and aspiration constraints. By integrating concave-envelope machinery with martingale duality, we derive explicit closed-form solutions for nonconcave utilities in complete and incomplete markets. Our results highlight that inverse S-shaped distortion induces excessive risk-taking and time inconsistency, necessitating pre-commitment strategies to align ex ante and interim optimality.

Future research may proceed three key extensions. First, developing robust optimization frameworks to address ambiguity in behavioral parameters and model misspecification is crucial for practical stability. Second, the model may be extended to endogenize policyholder behavior, particularly lapse risks and surrender options, requiring a game-theoretic approach to capture strategic interactions between asset management and liability decisions; see, e.g., Stackelberg games in \cite{CS2019}. Third, empirical methods can help calibrate distortion parameters from institutional portfolio data, validating the predicted regime switches. Addressing these challenges will further integrate behavioral finance into actuarial science, offering more resilient tools for balance sheet management.



%

\quad

\noindent
{\bf Acknowledgements.}
Y. Liu acknowledges financial support from the National Natural Science Foundation of China (Grant No. 12401624), The Chinese University of Hong Kong (Shenzhen) university development fund (Grant No. UDF01003336) and Shenzhen Science and Technology Program (Grant Nos. RCBS20231211090814028, JCYJ20250604141203005, 2025TC0010) and is partly supported by the Guangdong Provincial Key Laboratory of Mathematical Foundations for Artificial Intelligence (Grant No. 2023B1212010001).  
The authors are grateful to Jose Blanchet for helpful discussions and to members of the research group on financial mathematics and risk management at The Chinese University of Hong Kong, Shenzhen for their useful feedback and conversations.


\appendix

\section{Proofs of results in Section \ref{formula}}\label{proof}
\subsection{Proof of Proposition \ref{prop:optimal wealth process}}
Consider the following functions:
\begin{equation}\label{eq:optimal quantile}
\begin{aligned}
    Q_{L}(\lambda,z) &:= I_{L}\(\lambda \hat{\varphi}'\(z\) \)\id_{\{z \leq 1-w(\alpha) \} } + \[I_{L}\(\lambda \hat{\varphi}'\(z\) \) \vee L \]\id_{\{z > 1-w(\alpha) \} },\\
    Q_{R}(\lambda,z) &:= I_{R}\(\lambda \hat{\varphi}'\(z\) \)\id_{\{z \leq 1-w(\alpha) \} } + \[I_{R}\(\lambda \hat{\varphi}'\(z\) \) \vee L \]\id_{\{z > 1-w(\alpha) \} },
\end{aligned}
\end{equation}
where we define
\begin{equation*}
\begin{aligned}
    I_{L}(y) &:= \inf\{x \in \D: (U^{**})'(x) \leq y \},\\
    I_{R}(y) &:= \sup\{x \in \D:(U^{**})'(x) \geq y \}.
\end{aligned}
\end{equation*}

We proceed to introduce four optimization problems.
\begin{itemize}
    \item \textbf{Problem A}: 
    $\begin{aligned}
\text{ }V_A(x_0) = \sup_{Q \in \mathcal{G}_L} \int_{0}^{1} U(Q(x))\d x,
\textrm{ subject to }    \int_0^1 Q(x)\varphi'(x) dx = x_0+H(0).\\
\end{aligned}$
    \item \textbf{Problem B}: 
    $\begin{aligned}
\text{ }V_B(\lambda,x_0) = \sup_{Q \in \mathcal{G}_L} \int_{0}^{1}\left( U(Q(x)) - 
\lambda \varphi'(x)Q(x) \right)\d x + \lambda (x_0+H(0)).\\
\end{aligned}$
    \item \textbf{Problem C}: $\begin{aligned}
\text{ }V_C(\lambda,x_0) = \sup_{Q \in \mathcal{G}_L} \int_{0}^{1}\left( \hat{U}(Q(x)) - 
\lambda \varphi'(x)Q(x) \right)\d x + \lambda (x_0+H(0)).\\
\end{aligned}$
    \item \textbf{Problem D}: $\begin{aligned}
\text{ }V_D(\lambda,x_0) = \sup_{Q \in \mathcal{G}_L} \int_{0}^{1}\left( \hat{U}(Q(x)) - 
\lambda \hat{\varphi}'(x)Q(x) \right)\d x + \lambda (x_0+H(0)).\\
\end{aligned}$
\end{itemize}
Theorem 3.1 in \cite{Bi2023} shows a sufficient condition for Problems (A)--(D) to be equivalent without the aspiration constraint. We verify that the equivalence also holds with the constraint. 

We start with Problem (D). Here, we follow the solution of Problem (5.1) in \cite{HZ2016} to solve this problem by pointwise maximization of the integrand. To simplify the notation, we denote 
\begin{equation*}
    h(x,q) := \hat{U}\(q\) - \lambda \hat{\varphi}'(x) q,\quad 0 \leq x \leq 1. 
\end{equation*}
Then, the objective function is 
\begin{equation*}
    \int_0^1 h(x, Q(x))\d x + \lambda (x_0+H(0)).
\end{equation*}
For each $x \in (0,1-w(\alpha)]$, it is clear that both $q_L^* = I_L(\lambda \hat{\varphi}'(x))$ and $q_R^* = I_R(\lambda\hat{\varphi}'(x) ) $ are maximizers. When $x \in (1-w(\alpha),1)$, we find both $q_L^* = I_L(\lambda \hat{\varphi}'(x))\vee L$ and $q_R^* = I_R(\lambda \hat{\varphi}'(x))\vee L$ are maximizers. Hence, we can introduce the following two solutions to Problem (D):
\begin{equation*}
\begin{aligned}
    Q_L(\lambda, z) &= I_L\(\lambda \hat{\varphi}'(z) \)\id_{\{z\leq 1-w(\alpha) \}} + \[I_L\(\lambda \hat{\varphi}'(z)\) \vee L \]\id_{\{z > 1-w(\alpha)\} },\\
    Q_R(\lambda, z) &= I_R\(\lambda \hat{\varphi}'(z) \)\id_{\{z\leq 1-w(\alpha) \}} + \[I_R\(\lambda \hat{\varphi}'(z)\) \vee L \]\id_{\{z > 1-w(\alpha)\} }.
\end{aligned}
\end{equation*}

We have shown the solutions to Problem (D). From Theorem 3.1 in \cite{Bi2023}, we see that if $L \notin \{x \in \R:U^{**} (x) > U(x) \}$, then $Q_L(\lambda,\cdot)$ and $Q_R(\lambda, \cdot)$ are solutions to Problem (B). Recall the definition of $\Lambda_n$. 
From Theorem 4.2 in \cite{Bi2023}, we have the following results:
\begin{enumerate}[(i)]
    \item if $x_0+H(0) \in \(\cup_{k=1}^n I_k\)^C \cup \(A \exp\left(\mu_{\boldsymbol{\theta}} + \frac{\sigma_{\boldsymbol{\theta}}^2}{2}\right)\Phi^{-1}\left(\beta - \sigma_{\boldsymbol{\theta}}\right), +\infty\)$, then $Q_L(\lambda^*,x_0)$ is the unique solution to Problem (A), where $\lambda^*$ is the unique solution to $\int_0^1 Q_L(\lambda,x)\varphi'(x)\d x = x_0+H(0)$;
    \item if $x_0+H(0)= x_{L,k}$ for some $k$, then $Q_L(\lambda_k,\cdot)$ is the unique solution to Problem (A);
    \item if $x_0+H(0) = x_{R,k}$ for some $k$, then $Q_R(\lambda_k,\cdot)$ is the unique solution to Problem (A);
    \item otherwise, then our method fails to find the optimal solution.
\end{enumerate}


By the change-of-variable method, the optimal terminal wealth is given by
\begin{equation*}
    Y_T^* = G^*\(1 - F_{\xi_{\btheta}}\(\xi_{\btheta} \) \) = Q_{\lambda^*}^* \(1 - w\(F_{\xi_{\btheta}}\(\xi_{\btheta}\) \) \).
\end{equation*}
We can replicate this terminal wealth by defining
\begin{equation*}
    Y_t^* := \xi_t^{-1} \E \[\xi_T Y_T^*|\F_t \].
\end{equation*}
Based on the initial wealth, there exist four cases of the optimal wealth process:

\noindent
Case 1. When $x_0+H(0) \in \(\cup_{k=1}^n I_k\)^C \cup \(A \exp\left(\mu_{\boldsymbol{\theta}} + \frac{\sigma_{\boldsymbol{\theta}}^2}{2}\right)\Phi^{-1}\left(\beta - \sigma_{\boldsymbol{\theta}}\right), +\infty\)$, we have
\begin{equation*}
\begin{aligned}
    X_T^*&= \[Y_T^*\vee L \]\id_{\left\{\xi_T \leq F_{\xi_{\btheta}}^{-1}\(\alpha\)\right\}} + Y_T^*\id_{\left\{\xi_T > F_{\xi_{\btheta}}^{-1}\(\alpha\) \right\}}.
\end{aligned}
\end{equation*}
Then we only need to compute the following conditional expectation to obtain the explicit form of the optimal wealth process:
\begin{equation*}
    X_t^* = \xi_t^{-1}\E\[\xi_T X_T|\F_t\].
\end{equation*}
We first compute several conditional expectations that will be used repeatedly. For any $0\leq v<u\leq 1$, we have
\begin{equation*}
\left\{v<1-w(F_{\xi_{\btheta}}(\xi_{\btheta}))<u\right\} 
= \left\{F_{\xi_{\btheta}}^{-1}\left(w^{-1}(1-u)\right)<\xi_{\btheta}
<F_{\xi_{\btheta}}^{-1}\left(w^{-1}(1-v)\right)\right\}.
\end{equation*}
Moreover, recall that
\begin{equation*}
\xi_{\btheta}= \xi_t \exp\left\{-\int_t^T\left(r_s+\frac{\|\btheta_s\|_2^2}{2}\right)\d s-N\sqrt{\int_t^T\|\btheta_s\|_2^2\d s}\right\}.
\end{equation*}
By the definition of $\hat d(t,y)$, we have
\begin{equation*}
\left\{v<1-w(F_{\xi_{\btheta}}(\xi_{\btheta}))<u\right\} =\left\{\hat d(t,v)<N<\hat d(t,u)\right\}.
\end{equation*}
Therefore,
\begin{equation}\label{eq:basic-conditional-expectation}
\begin{aligned}
&\quad\xi_t^{-1}\mathbb E\left[\xi_{\btheta}\id_{\{v<1-w(F_{\xi_{\btheta}}(\xi_{\btheta}))<u\}}\Big|\F_t\right] \\
&=e^{-\int_t^T r_s\d s}\int_{\hat d(t,v)}^{\hat d(t,u)}\exp\left\{-z\sqrt{\int_t^T\|\btheta_s\|_2^2\d s}-\frac12\int_t^T\|\btheta_s\|_2^2\d s\right\}\Phi'(z)\d z \\
&=e^{-\int_t^T r_s\d s}\int_{d(t,v)}^{d(t,u)}\Phi'(z)\d z\\
&=e^{-\int_t^T r_s\d s}\left[\Phi(d(t,u))-\Phi(d(t,v))\right].
\end{aligned}
\end{equation}
Now, we can apply \eqref{eq:basic-conditional-expectation} to the three local forms of the PHARA utility. We begin with the linear part. On the event
\begin{equation*}
\left\{\lambda\hat{\varphi}'\left(1-w(F_{\xi_{\btheta}}(\xi_{\btheta}))\right)\in(\gamma_k^+,\gamma_k^-)\right\},
\end{equation*}
the pointwise optimizer equals $a_k$. Since $\hat{\varphi}'$ is non-increasing
and $I$ is the generalized inverse, this event is equivalent to
\begin{equation*}
\left\{I\left(\frac{\gamma_k^-}{\lambda}\right)<1-w(F_{\xi_{\btheta}}(\xi_{\btheta}))<I\left(\frac{\gamma_k^+}{\lambda}\right)\right\}.
\end{equation*}
Hence, by \eqref{eq:basic-conditional-expectation}, we have
\begin{equation}\label{eq:linear-expectation-proof}
\begin{aligned}
&\quad\xi_t^{-1}\mathbb E\left[\xi_{\btheta}a_k\id_{\left\{\lambda\hat{\varphi}'\left(1-w(F_{\xi_{\btheta}}(\xi_{\btheta}))\right)\in(\gamma_k^+,\gamma_k^-)\right\}}\mid\mathcal F_t\right] \\
&=e^{-\int_t^T r_s\d s}a_k\left[\Phi\left(d\left(t,I\left(\frac{\gamma_k^+}{\lambda}\right)\right)\right)-\Phi\left(d\left(t,I\left(\frac{\gamma_k^-}{\lambda}\right)\right)\right)\right].
\end{aligned}
\end{equation}
After incorporating the aspiration constraint, the difference of distribution functions in \eqref{eq:linear-expectation-proof} is replaced by $P$. Therefore,
\begin{equation}\label{eq:XD-conditional-proof}
X^D_{t,k,\lambda}=e^{-\int_t^T r_s\d s}a_kP\left(t,\lambda,I\left(\frac{\gamma_k^+}{\lambda}\right),I\left(\frac{\gamma_k^-}{\lambda}\right)\right).
\end{equation}
Next, consider the HARA part where $R_k\notin\{0,\infty\}$. On the event
\begin{equation*}
\lambda\hat{\varphi}'\left(1-w(F_{\xi_{\btheta}}(\xi_{\btheta}))\right)\in(\gamma_{k+1}^-,\gamma_k^+),
\end{equation*}
the pointwise optimizer equals
\begin{equation*}
A_k+\left(\frac{\gamma_k^+}{\lambda\hat{\varphi}'\left(1-w(F_{\xi_{\btheta}}(\xi_{\btheta}))\right)}\right)^{R_k^{-1}}(a_k-A_k).
\end{equation*}
We still have the equivalence:
\begin{equation}\label{eq:power-event-proof}
\left\{\lambda\hat{\varphi}'\left(1-w(F_{\xi_{\btheta}}(\xi_{\btheta}))\right)\in(\gamma_{k+1}^-,\gamma_k^+)\right\} =\left\{I\left(\frac{\gamma_k^+}{\lambda}\right)<1-w(F_{\xi_{\btheta}}(\xi_{\btheta}))<I\left(\frac{\gamma_{k+1}^-}{\lambda}\right)\right\}.
\end{equation}
The constant term $A_k$ is evaluated directly by \eqref{eq:basic-conditional-expectation}. For the power term, the explicit form of $\hat{\varphi}'$ gives
\begin{equation}\label{eq:power-transform-proof}
\left(\frac{\gamma_k^+}{\lambda\hat{\varphi}'\left(1-w(F_{\xi_{\btheta}}(\xi_{\btheta}))\right)}\right)^{R_k^{-1}}=\left(\frac{\tilde{k}\gamma_k^+}{\lambda}\right)^{R_k^{-1}}\exp\left\{\frac{a\mu_{\btheta}}{R_k\sigma_{\btheta}}\right\}\xi_{\btheta}^{-c_k}.
\end{equation}
Thus, for any $0\leq v<u\leq1$, we have
\begin{equation}\label{eq:power-expectation-core-proof}
\begin{aligned}
&\quad\xi_t^{-1}\mathbb E\left[\xi_{\btheta}^{1-c_k}\id_{\{v<1-w(F_{\xi_{\btheta}}(\xi_{\btheta}))<u\}}\Big|\F_t\right] \\
&=\xi_t^{-c_k}\exp\left\{-(1-c_k)\int_t^T r_s\d s-\frac{1-c_k}{2}\int_t^T\|\btheta_s\|_2^2\d s\right\}\int_{\hat d(t,v)}^{\hat d(t,u)}\exp\left\{-(1-c_k)z\sqrt{\int_t^T\|\btheta_s\|_2^2\d s}\right\}\Phi'(z)\d z \\
&=e^{-\int_t^T r_s\d s}\xi_t^{-c_k}\exp\left\{c_k\int_t^T r_s\d s+\frac{c_k(c_k-1)}{2}\int_t^T\|\btheta_s\|_2^2\d s\right\}\left[\Phi(d_k(t,u))-\Phi(d_k(t,v))\right].
\end{aligned}
\end{equation}
Combining \eqref{eq:power-transform-proof} and \eqref{eq:power-expectation-core-proof} and taking $u=I\left(\gamma_{k+1}^-/\lambda\right)$ and $v=I\left(\gamma_k^+/\lambda\right)$, we obtain
\begin{equation}\label{eq:XP-conditional-proof}
\begin{aligned}
X_{t,k,\lambda}^P &:= e^{-\int_t^T r_s \,\d s}\left\{A_k P\(t,\lambda, I\(\frac{\gamma_{k+1}^-}{\lambda} \), I\(\frac{\gamma_k^+}{\lambda} \) \) \right.\nonumber\\
    &\quad \left.+\(a_{k}-A_{k} \)B_{t,k}\(\frac{\tilde{k}\gamma_{k}^+}{\lambda} \)^{R_{k}^{-1}} \xi_t^{-c_{k}}
    P_k\(t,\lambda,I\(\frac{\gamma_{k+1}^-}{\lambda} \), I\(\frac{\gamma_k^+}{\lambda} \) \) \right\}\id_{\{R_k \neq 0, \infty \}}.
\end{aligned}
\end{equation}
Now we consider the exponential part where $R_k=\infty$, $A_k=-\infty$, and $\alpha_k>0$. On the same event as in \eqref{eq:power-event-proof}, the point-wise optimizer equals
\begin{equation*}
a_k+\frac1{\alpha_k}\log\left(\frac{\gamma_k^+}{\lambda\hat{\varphi}'\left(1-w(F_{\xi_{\btheta}}(\xi_{\btheta}))\right)}\right).
\end{equation*}
The constant term $a_k$ is evaluated by \eqref{eq:basic-conditional-expectation}. For the logarithmic term, we use
\begin{equation*}
\log\left(\frac{\gamma_k^+}{\lambda\hat{\varphi}'\left(1-w(F_{\xi_{\btheta}}(\xi_{\btheta}))\right)}\right) = \log\left(\frac{\tilde{k}\gamma_k^+}{\lambda}\right)+\frac{a\mu_{\btheta}}{\sigma_{\btheta}}-\left(1+\frac a{\sigma_{\btheta}}\right)\log\xi_{\btheta}.
\end{equation*}
Moreover,
\begin{equation*}
\log\(\xi_{\btheta}\)=\log\(\xi_t\)-\int_t^T\left(r_s+\frac{\|\btheta_s\|_2^2}{2}\right)\d s-N\sqrt{\int_t^T\|\btheta_s\|_2^2\d s}.
\end{equation*}
It follows that, for $0\leq v<u\leq1$,

\begin{equation}\label{eq:log-expectation-proof}
\begin{aligned}
&\quad\xi_t^{-1}\mathbb E\left[\xi_{\btheta}\log\left(\frac{\gamma_k^+}{\lambda\hat{\varphi}'\left(1-w(F_{\xi_{\btheta}}(\xi_{\btheta}))\right)}\right)\id_{\{v<1-w(F_{\xi_{\btheta}}(\xi_{\btheta}))<u\}}\mid\mathcal F_t\right] \\
&=e^{-\int_t^T r_s\d s}\Bigg\{\Bigg[\log\left(\frac{\tilde{k}\gamma_k^+}{\lambda\xi_t^{1+\frac a{\sigma_{\btheta}}}}\right)+\frac{a\mu_{\btheta}}{\sigma_{\btheta}}+\left(1+\frac a{\sigma_{\btheta}}\right)\int_t^T\left(r_s-\frac{\|\btheta_s\|_2^2}{2}\right)\d s\Bigg]\left[\Phi(d(t,u))-\Phi(d(t,v))\right] \\
&\quad-\left(1+\frac a{\sigma_{\btheta}}\right)\sqrt{\int_t^T\|\btheta_s\|_2^2\d s}\left[\Phi'(d(t,u))-\Phi'(d(t,v))\right]\Bigg\}.
\end{aligned}
\end{equation}
After the aspiration adjustment, the exponential term is
\begin{align}\label{eq:XE-conditional-proof}
X_{t,k,\lambda}^E 
    &=e^{-\int_t^T r_s \,\d s}\left\{a_k P\(t,\lambda, I\(\frac{\gamma_{k+1}^-}{\lambda}\), I\(\frac{\gamma_k^+}{\lambda} \) \) \right.\nonumber\\
    &\quad\quad + \frac{1}{\alpha_k}\left\{\(\log\( \frac{\tilde{k} \gamma_{k}^+ }{\lambda \xi_t^{1+\frac{a}{\sigmatheta}}} \)+\frac{a\mutheta }{\sigmatheta} \)P\(t,\lambda,I\(\frac{\gamma_{k+1}^-}{\lambda}\),I\(\frac{\gamma_k^+}{\lambda}\) \) \right.\nonumber\\
    &\quad\quad\quad  +\(1+\frac{a}{\sigmatheta}\)\Bigg(\int_t^T\(r_s-\frac{\|\btheta_s\|_2^2}{2} \)\,\d s P\(t,\lambda,I\(\frac{\gamma_{k+1}^-}{\lambda}\), I\(\frac{\gamma_k^+}{\lambda} \) \) \nonumber\\
    &\quad\quad\quad\quad\left.\left.\left. -\sqrt{\int_t^T\|\btheta_s\|_2^2\,\d s} Q\(t,\lambda,I\(\frac{\gamma_{k+1}^-}{\lambda}\), I\(\frac{\gamma_k^+}{\lambda}\) \) \) \right\}\right\}\id_{\left\{R_{k}=\infty \right\} }.
\end{align}
Finally, the aspiration constraint contributes an additional term. Since $Y^*<L$ is equivalent to
\begin{equation*}
1-w(F_{\xi_{\btheta}}(\xi_{\btheta}))<I\left(\frac{U'(L)}{\lambda}\right),
\end{equation*}
and the benchmark region is
\begin{equation*}
    1-w(F_{\xi_{\btheta}}(\xi_{\btheta}))>1-w(\alpha),
\end{equation*}
the range is restricted as
\begin{equation*}
\left\{1-w(\alpha)<1-w(F_{\xi_{\btheta}}(\xi_{\btheta}))<I\left(\frac{U'(L)}{\lambda}\right)\right\}.
\end{equation*}
Thus, by \eqref{eq:basic-conditional-expectation}, we have 
\begin{equation}\label{eq:floor-conditional-proof}
\begin{aligned}
&\quad\xi_t^{-1}\mathbb E\left[\xi_{\btheta}L\id_{\left\{1-w(\alpha)<1-w(F_{\xi_{\btheta}}(\xi_{\btheta}))<I\left(\frac{U'(L)}{\lambda}\right)\right\}}\mid\mathcal F_t\right] \\
&=e^{-\int_t^T r_s\d s}L\left[\Phi\left(d\left(t,I\left(\frac{U'(L)}{\lambda}\right)\right)\right)-\Phi(d(t,1-w(\alpha)))\right]\id_{\left\{1-w(\alpha)<I\left(\frac{U'(L)}{\lambda}\right)\right\}}.
\end{aligned}
\end{equation}
Combining \eqref{eq:XD-conditional-proof}, \eqref{eq:XP-conditional-proof},
\eqref{eq:XE-conditional-proof} and \eqref{eq:floor-conditional-proof} yields
the complete form of \eqref{eq:wealth process basic}.

Based on the initial wealth level, there are four cases for the optimal wealth process. We emphasize that the affine part of $\hat{\varphi}$ on $(0,c)$ must be treated separately. Indeed, since $\hat{\varphi}$ is affine on $(0,c)$, we
have
\begin{equation*}
    \hat{\varphi}'(z)=\varphi'(c),\quad z\in(0,c).
\end{equation*}
Consequently, on $(0,c)$, the point-wise optimizer is determined by the single value $\lambda\varphi'(c)$. This produces the auxiliary terms in the wealth process.

\noindent
Case 1. Suppose that $x_0+H(0)\notin\bigcup_{k=1}^n I_k$, and let $\lambda^*$ be the unique Lagrange multiplier satisfying
\begin{equation*}
    \int_0^1 Q_L(\lambda^*,x)\varphi'(x)\d x=x_0+H(0).
\end{equation*}
Then
\begin{equation*}
Y_T^* =Q_L\left(\lambda^*,1-w(F_{\xi_{\btheta}}(\xi_{\btheta}))\right).
\end{equation*}
We further distinguish the following subcases according to the location of $\lambda^*\varphi'(c)$.

\noindent
Case 1(i). If $\lambda^*\varphi'(c)>\gamma_0^+$, then the affine part $(0,c)$ of $\hat{\varphi}$ does not generate any auxiliary term. Therefore,
\begin{equation*}
\begin{aligned}
X_t^*&=-H(t)+\sum_{k=0}^n\(X^D_{t,k,\lambda^*}+X^P_{t,k,\lambda^*}+X^E_{t,k,\lambda^*}\) \\
&\quad+e^{-\int_t^T r_s\d s}L\left[\Phi\left(d\left(t,I\left(\frac{U'(L)}{\lambda^*}\right)\right)\right)-\Phi(d(t,1-w(\alpha)))\right]\id_{\left\{1-w(\alpha)<I\left(\frac{U'(L)}{\lambda^*}\right)\right\}}.
\end{aligned}
\end{equation*}

\noindent
Case 1(ii). Suppose that there exists $k^*\in\{0,\ldots,n-1\}$ such that $\gamma_{k^*}^+>\lambda^*\varphi'(c)>\gamma_{k^*+1}^+$. Since $\hat{\varphi}$ is affine on $(0,c)$, we have
$\hat{\varphi}'(z)=\varphi'(c)$ for $z\in(0,c)$. Hence the pointwise optimizer on $(0,c)$ is determined by $\lambda^*\varphi'(c)$.
When $R_{k^*}\notin\{0,\infty\}$, the value of the optimizer on $(0,c)$ is
\begin{equation*}
A_{k^*}+\left(\frac{\gamma_{k^*}^+}{\lambda^*\varphi'(c)}\right)^{R_{k^*}^{-1}}(a_{k^*}-A_{k^*}).
\end{equation*}
Therefore, the contribution from $(0,c)$ is
\begin{equation*}
\begin{aligned}
&e^{-\int_t^T r_s\d s}\left[A_{k^*}+\left(\frac{\gamma_{k^*}^+}{\lambda^*\varphi'(c)}\right)^{R_{k^*}^{-1}}(a_{k^*}-A_{k^*})\right]P(t,\lambda^*,c,0)\id_{\left\{c<I\left(\frac{\gamma_{k^*+1}^-}{\lambda^*}\right)\right\}}.
\end{aligned}
\end{equation*}
On the remaining part $\left(c,I\left(\gamma_{k^*+1}^-/\lambda^*\right)\right)$, the optimizer coincides with the ordinary $k^*$-th HARA segment. Applying the conditional expectation formula for the HARA part gives
\begin{equation*}
\begin{aligned}
&e^{-\int_t^T r_s\d s}\Bigg\{A_{k^*}P\left(t,\lambda^*,\max\left\{I\left(\frac{\gamma_{k^*+1}^-}{\lambda^*}\right),c\right\},c\right) \\
&\quad+(a_{k^*}-A_{k^*})B_{t,k^*}\left(\frac{\tilde{k}\gamma_{k^*}^+}{\lambda^*}\right)^{R_{k^*}^{-1}}\xi_t^{-c_{k^*}}P_{k^*}\left(t,\lambda^*,\max\left\{I\left(\frac{\gamma_{k^*+1}^-}{\lambda^*}\right),c\right\},c\right)\Bigg\}.
\end{aligned}
\end{equation*}
Combining these two pieces yields the auxiliary term in Case 1(ii) with $\id_{\{R_{k^*}\notin\{0,\infty\}\}}$.
On the other hand, when $R_{k^*}=\infty$, the value of the optimizer on $(0,c)$ is
\begin{equation*}
a_{k^*}+\frac1{\alpha_{k^*}}\log\left(\frac{\gamma_{k^*}^+}{\lambda^*\varphi'(c)}\right).
\end{equation*}
Thus the contribution from $(0,c)$ is
\begin{equation*}
\begin{aligned}
&e^{-\int_t^T r_s\d s}\left(a_{k^*}+\frac1{\alpha_{k^*}}\log\left(\frac{\gamma_{k^*}^+}{\lambda^*\varphi'(c)}\right)\right)P(t,\lambda^*,c,0).
\end{aligned}
\end{equation*}
On the remaining part $\left(c,I\left(\gamma_{k^*+1}^-/\lambda^*\right)\right)$,
the optimizer coincides with the ordinary exponential segment. Applying the conditional expectation formula for the exponential part gives
\begin{align}
&\quad e^{-\int_t^T r_s \,\d s}\left\{\(a_{k^{*}}+\frac{1}{\alpha_{k^{*}}}\log\(\frac{\gamma_{k^{*}}^+ }{\lambda^*\varphi'(c) } \) \)\Phi\(d(t,c) \)+a_{k^{*}}P\(t,\lambda^*,I\(\frac{\gamma_{k^*+1}^-}{\lambda^*}\),c  \)  \right\}\id_{\left\{R_{k^{*}}=\infty \right\} }\nonumber\\
    &\quad +\frac{1}{\alpha_{k^{*}}} e^{-\int_t^T r_s \,\d s}\left\{\(\log\( \frac{\tilde{k} \gamma_{k^{*}}^+ }{\lambda^* \xi_t^{1+\frac{a}{\sigmatheta}}} \)+\frac{a\mutheta }{\sigmatheta} \)P\(t,\lambda^*,I\(\frac{\gamma_{k^*+1}^-}{\lambda^*}\),c  \)\right.\nonumber\\
    &\quad\quad\; +\(1+\frac{a}{\sigmatheta}\)\(\int_t^T\(r_s-\frac{\|\btheta_s\|_2^2}{2} \)\,\d s P\(t,\lambda^*,I\(\frac{\gamma_{k^*+1}^-}{\lambda^*}\),c \) \right.\nonumber\\
    &\left.\left.\quad\quad\quad-\int_t^T\|\btheta_s\|_2^2\,\d sQ\(t,\lambda^*,I\(\frac{\gamma_{k^*+1}^-}{\lambda^*}\),c \) \) \right\}\id_{\left\{R_{k^{*}}=\infty \right\} }.\nonumber
\end{align}
Combining these two pieces yields the auxiliary term in Case 1(ii) with
$\id_{\{R_{k^*}=\infty\}}$. Together with the summation over $k=k^*+1,\ldots,n$ and the aspiration term, this proves the expression of $X_t^*$ in Case 1(ii).

\noindent
Case 2. If $x_0+H(0)=x_{L,k_0}$ for some $k_0$, then the left end-point is chosen on the affine part $(0,c)$. Thus the selected value on $(0,c)$ is $a_{k_0}$ and
\begin{equation*}
Y_T^*=Q_L\left(\lambda_{k_0},1-w(F_{\xi_{\btheta}}(\xi_{\btheta}))\right).
\end{equation*}
Consequently,
\begin{equation*}
\begin{aligned}
X_t^*&=-H(t)+e^{-\int_t^T r_s\d s}a_{k_0}P(t,\lambda_{k_0},c,0) \\
&\quad+\sum_{k=k_0+1}^n\left(X^D_{t,k,\lambda_{k_0}}+X^P_{t,k,\lambda_{k_0}}+X^E_{t,k,\lambda_{k_0}}\right) \\
&\quad+e^{-\int_t^T r_s\d s}L\left[\Phi\left(d\left(t,I\left(\frac{U'(L)}{\lambda_{k_0}}\right)\right)\right)-\Phi(d(t,1-w(\alpha)))\right]\id_{\left\{1-w(\alpha)<I\left(\frac{U'(L)}{\lambda_{k_0}}\right)\right\}}.
\end{aligned}
\end{equation*}

\noindent
Case 3. If $x_0+H(0)=x_{R,k_0}$ for some $k_0$, then the right end-point is chosen on the affine part $(0,c)$. Thus the selected value on $(0,c)$ is $a_{k_0+1}$ and
\begin{equation*}
Y_T^* = Q_R\left(\lambda_{k_0},1-w(F_{\xi_{\btheta}}(\xi_{\btheta}))\right).
\end{equation*}
Consequently,
\begin{equation*}
\begin{aligned}
X_t^*&=-H(t)+e^{-\int_t^T r_s\d s}a_{k_0+1}P(t,\lambda_{k_0},c,0) \\
&\quad+\sum_{k=k_0+1}^n\left(X^D_{t,k,\lambda_{k_0}}+X^P_{t,k,\lambda_{k_0}}+X^E_{t,k,\lambda_{k_0}}
\right) \\
&\quad+e^{-\int_t^T r_s\d s}L\left[\Phi\left(d\left(t,I\left(\frac{U'(L)}{\lambda_{k_0}}\right)\right)\right)-\Phi(d(t,1-w(\alpha)))\right]\id_{\left\{1-w(\alpha)<I\left(\frac{U'(L)}{\lambda_{k_0}}\right)\right\}}.
\end{aligned}
\end{equation*}

\noindent
Case 4. If $x_0+H(0)\in\bigcup_{k=1}^n I_k$, our method fails to find an optimal solution.

\subsection{Proof of Theorem \ref{thm:optimal portfolio}}


By Proposition \ref{prop:optimal wealth process}, for each fixed $t\in[0,T)$, the optimal wealth process $X_t^*$ can be regarded as a function of $\xi_t$. Since $H(t)$ is deterministic, we have
\begin{equation*}
 \frac{\partial Y_t^*(\xi_t)}{\partial \xi_t} = \frac{\partial X_t^*(\xi_t)}{\partial \xi_t}.
\end{equation*}
Applying Itô's formula to $\xi_tY_t^*$ and comparing the diffusion term with the self-financing wealth equation, we obtain
\begin{equation}\label{eq:feedback-pi-proof}
\boldsymbol{\pi}_t^*=-\left(\boldsymbol{\sigma}_t\boldsymbol{\sigma}_t^\top\right)^{-1}\left(\boldsymbol{\mu}_t-r_t\mathbf{1}_m\right)\xi_t\frac{\partial X_t^*(\xi_t)}{\partial \xi_t}.
\end{equation}
Thus it remains to differentiate the terms in Proposition \ref{prop:optimal wealth process} with respect to
$\xi_t$. By the definitions of $d$, $\hat d$, $d_k$ and $\hat d_k$, for any fixed $y\in[0,1]$, we have
\begin{equation}\label{eq:d-derivatives-proof}
\begin{aligned}
\xi_t\frac{\partial d(t,y)}{\partial \xi_t}&=\xi_t\frac{\partial \hat d(t,y)}{\partial \xi_t}=\frac{1}{\sqrt{\int_t^T\|\btheta_s\|_2^2\d s}},\\
\xi_t\frac{\partial d_k(t,y)}{\partial \xi_t}&=\xi_t\frac{\partial \hat d_k(t,y)}{\partial \xi_t}=\frac{1}{\sqrt{\int_t^T\|\btheta_s\|_2^2\d s}}.
\end{aligned}
\end{equation}
Therefore, for fixed $\lambda,u,v$, we have
\begin{equation}\label{eq:P-derivatives-proof}
\begin{aligned}
\xi_t\frac{\partial}{\partial \xi_t}P(t,\lambda,u,v)&=\frac{1}{\sqrt{\int_t^T\|\btheta_s\|_2^2\d s}}Q(t,\lambda,u,v),\\
\xi_t\frac{\partial}{\partial \xi_t}P_k(t,\lambda,u,v)&=\frac{1}{\sqrt{\int_t^T\|\btheta_s\|_2^2\d s}}Q_k(t,\lambda,u,v).
\end{aligned}
\end{equation}
Moreover, since $\Phi''(x)=-x\Phi'(x)$, for any function $D(\xi_t)$ appearing inside $Q(t,\lambda,u,v)$, we have
\begin{equation}\label{eq:Q-derivative-rule-proof}
\xi_t\frac{\partial}{\partial \xi_t}\Phi'(D(\xi_t)) = -\frac{D(\xi_t)}{\sqrt{\int_t^T\|\btheta_s\|_2^2\d s}}\Phi'(D(\xi_t)).
\end{equation}
The same rule applies to $Q_k(t,\lambda,u,v)$ with $d_k$ and $\hat d_k$. Remarkably, the possible switching points generated by $\vee$ and $\wedge$ are treated by one-sided derivatives and do not affect the resulting portfolio process.

We now differentiate the three types of terms. For the linear part, by \eqref{eq:P-derivatives-proof},
\begin{equation}\label{eq:XD-derivative-proof}
\begin{aligned}
\xi_t\frac{\partial X^D_{t,k,\lambda}}{\partial \xi_t}&=e^{-\int_t^T r_s\d s}\frac{a_k}{\sqrt{\int_t^T\|\btheta_s\|_2^2\d s}}Q\left(t,\lambda,I\left(\frac{\gamma_k^+}{\lambda}\right),I\left(\frac{\gamma_k^-}{\lambda}\right)\right).
\end{aligned}
\end{equation}
Substituting \eqref{eq:XD-derivative-proof} into \eqref{eq:feedback-pi-proof} gives $\boldsymbol{\pi}^D_{t,k,\lambda}$. For the HARA part where $R_k\notin\{0,\infty\}$, we use
\begin{equation}\label{eq:power-derivative-proof}
\begin{aligned}
&\quad\xi_t\frac{\partial}{\partial \xi_t}\left[\xi_t^{-c_k}P_k(t,\lambda,u,v)\right] \\
&=\xi_t^{-c_k}\left[-c_kP_k(t,\lambda,u,v)+\frac{1}{\sqrt{\int_t^T\|\btheta_s\|_2^2\d s}}Q_k(t,\lambda,u,v)\right].
\end{aligned}
\end{equation}
Taking $u=I\left(\gamma_{k+1}^-/\lambda\right)$ and $v=I\left(\gamma_k^+/\lambda\right)$ and differentiating $X^P_{t,k,\lambda}$ give $\boldsymbol{\pi}^P_{t,k,\lambda}$. For the exponential part, write the coefficient in $X^E_{t,k,\lambda}$ as
\begin{equation*}
\begin{aligned}
L_{t,k,\lambda}&:=\log\left(\frac{\tilde{k}\gamma_k^+}{\lambda\xi_t^{1+\frac a{\sigma_{\btheta}}}}\right)+\frac{a\mu_{\btheta}}{\sigma_{\btheta}}+\left(1+\frac a{\sigma_{\btheta}}\right)\int_t^T\left(r_s-\frac{\|\btheta_s\|_2^2}{2}\right)\d s.
\end{aligned}
\end{equation*}
Then, by direct differentiation,
\begin{equation}\label{eq:L-derivative-proof}
\xi_t\frac{\partial L_{t,k,\lambda}}{\partial \xi_t}=-\left(1+\frac a{\sigma_{\btheta}}\right).
\end{equation}
Using Proposition \ref{prop:optimal wealth process}, the exponential term has the form
\begin{equation*}
\begin{aligned}
X^E_{t,k,\lambda}&=e^{-\int_t^T r_s\d s}\Bigg\{a_kP(t,\lambda,u,v)+\frac1{\alpha_k}\left[L_{t,k,\lambda}P(t,\lambda,u,v)-\sqrt{\int_t^T\|\btheta_s\|_2^2\d s}\,Q(t,\lambda,u,v)\right]\Bigg\},
\end{aligned}
\end{equation*}
where $u=I\left(\gamma_{k+1}^-/\lambda\right)$ and $v=I\left(\gamma_k^+/\lambda\right)$.
Therefore, combining \eqref{eq:P-derivatives-proof}, \eqref{eq:Q-derivative-rule-proof} and \eqref{eq:L-derivative-proof}, we get 
\begin{equation}\label{eq:XE-derivative-proof}
\begin{aligned}
\xi_t\frac{\partial X^E_{t,k,\lambda}}{\partial \xi_t}&=e^{-\int_t^T r_s\d s}\Bigg\{\frac{a_k}{\sqrt{\int_t^T\|\btheta_s\|_2^2\d s}}Q(t,\lambda,u,v) \\
&\quad+\frac1{\alpha_k}\Bigg[\frac{L_{t,k,\lambda}}{\sqrt{\int_t^T\|\btheta_s\|_2^2\d s}}Q(t,\lambda,u,v)-\left(1+\frac a{\sigma_{\btheta}}\right)\Bigg(P(t,\lambda,u,v) \\
&\qquad-\sqrt{\int_t^T\|\btheta_s\|_2^2\d s}\,\xi_t\frac{\partial Q(t,\lambda,u,v)}{\partial \xi_t}\Bigg)\Bigg]\Bigg\}.
\end{aligned}
\end{equation}
Substituting \eqref{eq:XE-derivative-proof} into \eqref{eq:feedback-pi-proof} gives $\boldsymbol{\pi}^E_{t,k,\lambda}$. It remains to differentiate the aspiration term. By \eqref{eq:d-derivatives-proof},
\begin{equation*}
\begin{aligned}
&\quad\xi_t\frac{\partial}{\partial\xi_t}\Bigg\{e^{-\int_t^T r_s\d s}L\left[\Phi\left(d\left(t,I\left(\frac{U'(L)}{\lambda}\right)\right)\right)-\Phi(d(t,1-w(\alpha)))\right]\id_{\left\{1-w(\alpha)<I\left(\frac{U'(L)}{\lambda}\right)\right\}}\Bigg\} \\
&=e^{-\int_t^T r_s\d s}\frac{L}{\sqrt{\int_t^T\|\btheta_s\|_2^2\d s}}\left[\Phi'\left(d\left(t,I\left(\frac{U'(L)}{\lambda}\right)\right)\right)-\Phi'(d(t,1-w(\alpha)))\right]\id_{\left\{1-w(\alpha)<I\left(\frac{U'(L)}{\lambda}\right)\right\}}.
\end{aligned}
\end{equation*}
Finally, the auxiliary terms in Case 1(ii), Case 2 and Case 3 are differentiated by the same rules. The HARA auxiliary term follows from \eqref{eq:power-derivative-proof}, while the exponential auxiliary term follows from \eqref{eq:XE-derivative-proof} with $u=I\left(\gamma_{k^*+1}^-/\lambda^*\right)$ and $v=c$. Combining these derivatives with the feedback identity \eqref{eq:feedback-pi-proof} yields the stated optimal portfolio process in all cases. 

\subsection{Proof of Proposition \ref{prop:asymptotic}}

We fix $t\in(0,T)$ and conduct an asymptotic analysis with respect to $\xi_t$. Recall from Theorem \ref{thm:optimal portfolio} that
\begin{equation}\label{eq:feedback-prop2}
\boldsymbol{\pi}_t^*=-\left(\boldsymbol{\sigma}_t\boldsymbol{\sigma}_t^\top\right)^{-1}\left(\boldsymbol{\mu}_t-r_t\mathbf{1}_m\right)\xi_t\frac{\partial X_t^*(\xi_t)}{\partial \xi_t}.
\end{equation}
First, consider the case where $\xi_t\rightarrow0$. By the definition of $d(t,y)$, for any fixed $y\in(0,1)$, we have
\begin{equation*}
d(t,y)\to-\infty,\quad d_k(t,y)\to-\infty,\quad\Phi(d(t,y))\to0,\quad\Phi'(d(t,y))\rightarrow0.
\end{equation*}
Hence, the only dominating term is the power term. From the explicit expression of $X_t^*$ in Proposition \ref{prop:optimal wealth process}, we obtain
\begin{equation*}
X_t^*\sim K_t\xi_t^{-\frac{a+\sigma_{\btheta}}{p\sigma_{\btheta}}},\quad \xi_t\rightarrow0,
\end{equation*}
where $K_t>0$ is independent of $\xi_t$. Consequently,
\begin{equation}\label{eq:good-state-derivative-prop2}
\xi_t\frac{\partial X_t^*(\xi_t)}{\partial \xi_t}\sim-\frac{a+\sigma_{\btheta}}{p\sigma_{\btheta}}X_t^*,\quad\xi_t\rightarrow0.
\end{equation}
Combining \eqref{eq:feedback-prop2} and \eqref{eq:good-state-derivative-prop2}, we have
\begin{equation}\label{eq:good-state-pi-ratio-prop2}
\frac{\boldsymbol{\pi}_t^*}{X_t^*}\rightarrow\frac{a+\sigma_{\btheta}}{p\sigma_{\btheta}}\left(\boldsymbol{\sigma}_t\boldsymbol{\sigma}_t^\top\right)^{-1}\left(\boldsymbol{\mu}_t-r_t\mathbf{1}_m\right),\quad\xi_t\rightarrow0.
\end{equation}
Since $X_t^*\rightarrow\infty$, it follows from \eqref{eq:good-state-pi-ratio-prop2} that $\boldsymbol{\pi}_t^*\rightarrow\infty$.

Next, consider the case where $\xi_t\rightarrow\infty$. By the definition of $d(t,y)$, for any fixed $y\in(0,1)$, we have
\begin{equation}\label{eq:bad-state-d-limit-prop2}
d(t,y)\rightarrow\infty,\quad d_k(t,y)\rightarrow\infty,\quad\Phi(d(t,y))\rightarrow1,\quad\Phi'(d(t,y))\rightarrow0.
\end{equation}
Since $\hat{\varphi}$ is affine on $(0,c)$, the limiting point-wise optimizer is determined by $\lambda\varphi'(c)$. Now we consider two cases. If $\lambda\varphi'(c)\leq \gamma^+$, then the point-wise optimizer on $(0,c)$ converges to $\frac{(1-\delta)L_T^g}{1-\delta\eta}$.
Using \eqref{eq:bad-state-d-limit-prop2} in the explicit formula of $X_t^*$, we get
\begin{equation*}
X_t^*\rightarrow\frac{(1-\delta)L_T^g}{1-\delta\eta},\quad\xi_t\frac{\partial X_t^*(\xi_t)}{\partial \xi_t}\rightarrow0.
\end{equation*}
Therefore, by \eqref{eq:feedback-prop2}, we have
\begin{equation*}
\boldsymbol{\pi}_t^*\rightarrow \mathbf{0},\qquad \frac{\boldsymbol{\pi}_t^*}{X_t^*}\rightarrow \mathbf{0}.
\end{equation*}
On the other hand, if $\lambda\varphi'(c)>\gamma^+$, from
\eqref{eq:bad-state-d-limit-prop2}, we have
\begin{equation*}
    X_t^*\rightarrow 0,\quad \xi_t\frac{\partial X_t^*(\xi_t)}{\partial \xi_t}\rightarrow0,\quad\boldsymbol{\pi}_t^*\rightarrow \mathbf{0}.
\end{equation*}
Since $d(t,c)\rightarrow\infty$, the wealth $X_t^*$ is of the same order as $1-\Phi(d(t,c))$, while $\xi_t\partial X_t^*/\partial\xi_t$ is of the same order as $\Phi'(d(t,c))$. Therefore,
\begin{equation}\label{eq:bad-state-case2-ratio-prop2}
-\frac{\xi_t\frac{\partial X_t^*(\xi_t)}{\partial \xi_t}}{X_t^*}\rightarrow \boldsymbol{\infty}.
\end{equation}
Combining \eqref{eq:feedback-prop2} and \eqref{eq:bad-state-case2-ratio-prop2}, we obtain
\begin{equation*}
\frac{\boldsymbol{\pi}_t^*}{X_t^*}=\left(\boldsymbol{\sigma}_t\boldsymbol{\sigma}_t^\top\right)^{-1}\left(\boldsymbol{\mu}_t-r_t\mathbf{1}_m\right)\left(-\frac{\xi_t\frac{\partial X_t^*(\xi_t)}{\partial \xi_t}}{X_t^*}\right)\rightarrow\boldsymbol{\infty}.
\end{equation*}
\end{document}